\documentclass[%
 reprint,
superscriptaddress,
 amsmath,amssymb,
 aps,
prb,
floatfix,
]{revtex4-2}

\usepackage{graphicx}
\usepackage{dcolumn}
\usepackage{amsmath,bm}
\usepackage{hyperref}
\usepackage{comment}

\usepackage{feynmp}
\usepackage{tikz-feynman}
\tikzfeynmanset{compat=1.1.0}

\usepackage[caption=false]{subfig}

\DeclareMathOperator{\Tr}{Tr}

\begin{document}

\preprint{APS/123-QED}

\title{Non-linear Diffusion processes and hydrodynamic cascades in lattice gas models}
\title{Diffusion cascade in a model of interacting random walkers}


\author{Abhishek Raj}
\email{abhishek654r@gmail.com}
\affiliation{Physics Program and Initiative for the Theoretical Sciences, The Graduate Center, CUNY, New York, New York 10016, USA}
 \affiliation{Department of Physics and Astronomy, College of Staten Island, CUNY, Staten Island, New York 10314, USA
 }

\author{Paolo Glorioso}
\affiliation{
 Zyphra, Palo Alto, CA 94306, USA\\
}%

\author{Sarang Gopalakrishnan}%
\affiliation{%
 Department of Electrical and Computer Engineering,
Princeton University, Princeton, NJ 08544, USA\\
}%

\author{Vadim Oganesyan}
\email{vadim.oganesyan@csi.cuny.edu}
 \affiliation{Physics Program and Initiative for the Theoretical Sciences, The Graduate Center, CUNY, New York, New York 10016, USA}
 \affiliation{Department of Physics and Astronomy, College of Staten Island, CUNY, Staten Island, New York 10314, USA
 }

\date{\today}

\begin{abstract}

We consider the relaxation of finite-wavevector density waves in a facilitated classical lattice gas. Linear hydrodynamics predicts that such perturbations should relax exponentially, but nonlinear effects were predicted to cause subexponential relaxation via nonperturbative long-time tails~\cite{heavyluca}. We present a detailed numerical study of this effect. While our results clearly indicate the importance of nonlinear effects, we find that the wavevector-dependence of the late-time relaxation is clearly inconsistent with theoretical predictions. We discuss manifestations of hydrodynamic nonlinearities in mesoscopic samples and at short times.




\end{abstract}

\maketitle


\section{Introduction}

The transport of conserved quantities (like charge or energy) in generic lattice models is diffusive: the lattice causes momentum relaxation (e.g., via Umklapp processes), so there are no long-lived currents. Diffusion has been familiar for centuries, but it was realized only in the 1970s that the linear diffusion equation gives qualitatively incorrect predictions for the late-time asymptotics of natural physical observables (such as the current-current correlator). These discrepancies from the linear theory are \emph{prima facie} surprising since all allowed nonlinearities (such as the density-dependence of the diffusion constant) are irrelevant in the renormalization group sense at the diffusive fixed point. In fact, nonlinearities are \emph{dangerously irrelevant}, and their effects have been catalogued under the rubric of ``long-time tails.'' Most familiar examples of long-time tails lead to power-law relaxation of physical observables, with exponents that can be read off from the diffusive fixed point; formally, they can be understood in terms of low-order perturbation theory in the nonlinearity. A few years ago, Delacretaz~\cite{heavyluca} found a somewhat different type of nonlinear effect: instead of arising at low orders in perturbation theory, it is essentially nonperturbative. This nonperturbative effect is of interest not only in the context of diffusive hydrodynamics but also much more generally in field theory: it is a simple example where ``heavy operators'' (i.e., operators that strongly perturb the vacuum) are manifestly important. (Heavy operators are a topic of widespread recent interest in the context of thermalization in CFTs~\cite{collier2020universal, karlsson2021thermalization}.) 

In the present work we explore this effect by performing numerical simulations on a facilitated classical lattice gas model. Before turning to our results, let us briefly summarize the argument of Ref.~\cite{heavyluca}. Consider perturbing an equilibrium high-temperature state by imprinting a weak density modulation at wavevector $Q$. In linear hydrodynamics this will decay as $\exp(-D Q^2 t)$. However, nonlinearities will mix a single density wave at $Q$ with a pair of density waves at (say) $Q/2$. The state with two density waves at $Q/2$ relaxes at the rate $\exp(-D Q^2 t / 2)$, so at late times it dominates over the single density wave at $Q$. We can apply this argument repeatedly to conclude that a single weak perturbation at $Q$ is likeliest to survive at very late times if it turns into a ``cascade'' of $n$ excitations each at wavevector $\sim Q/n$, where $n \to \infty$ as $t \to \infty$. Ref.~\cite{heavyluca} gave a heuristic diagram-counting argument to the effect that the late-time decay should scale as $\exp(-\sqrt{D Q^2 t})$; the reasoning, though plausible, rested on uncontrolled assumptions about the contributions of factorially many diagrams. Our objective is to test this prediction against explicit numerics on a model that is chosen to have strong nonlinearities. Our main conclusion is that, surprisingly, the time-dependence is consistent with Ref.~\cite{heavyluca} but the numerically extracted $Q$-dependence is not: instead, the late-time relaxation follows the form $f(Q) \exp(-\sqrt{C t})$, where $C$ is an empirically $Q$-independent scale that seems to be set by the lattice-scale physics.



Our basic conceptual tool is that of a weak quench, whereby the equilibrium state is excited initially with a single Fourier harmonic, e.g. for a density field $\rho(r)$
\begin{equation}
\rho(r,t=0)=\rho_{eq}+A \cos(Q r).
\end{equation}
Measuring operators $\mathcal{O}$ at later times gives a response function, which can be used (if we assume linear response) to infer \emph{equilibrium} two-time correlations \begin{equation}
    C_{\rho,\mathcal{O}}(Q,t)=\langle \mathcal{O}\rangle_Q/A,
\end{equation}
where $\langle\ldots\rangle_Q$ refers to the initial Fourier excited density matrix.  It is important that $A$ is small enough to avoid nonlinear corrections but large enough to give good signal-to-noise ratio. For lattice gas models defined in the next Section (and mostly studied near half-filling) we find empirically that there is linear-response behavior when $A\leq 0.2$.  Unless we explicitly want to discuss nonlinear effects (as in Section \ref{sec:2ndharm}), we assume (and spot check) linearity of the excitation and subsume the normalization by $A$ into the definitions of all observables discussed.

This paper is organized as follows. In Sec. \ref{sec:modelsmethods} we construct a family of stochastic lattice gas models with tunable nonlinearity. We compute the density-dependence of the diffusion constant and use this to deduce the value of the nonlinear hydrodynamic coupling; we then confirm this value by numerically studying the rate of second-harmonic generation. 
%
In Section \ref{sec:results} we present results of our numerical exploration of the diffusive cascade in several regimes using complementary stochastic and exact simulations at half-filling. We explore the two-time decay of Fourier modes of the density in the thermodynamic limit in Sec. \ref{sec:longtimes}, the spectrum of relaxation modes in mesoscopic systems in Sec: \ref{sec:meso}, and other signatures of the diffusion cascade in higher moments of wave amplitudes (Sec. \ref{sec:cascade}). We close in Sec. \ref{sec:summary} with a summary of results and discussion of open questions.


\section{Models and methods}
\label{sec:modelsmethods}
\subsection{Nonlinear random walk}

We consider a hardcore one-dimensional lattice gas with at most one particle per site. The lattice is updated by first subdividing into trimers (picking one of the three superlattices randomly at each update step) and then applying a random walk update to each trimer. Importantly, trimers with one and two particles per site can have different mobility.  Note, larger subdivisions can also work but trimers are the smallest subunits that allow for density dependent hopping. More explicitly, the transition rules we choose are: 1) The configuration 010 has probability $y/2$ of transforming into either 100 or 001, effectively allowing the particle to hop to an adjacent site, and has probability $1-y$ of staying 010. The reverse of these transitions have the same probabilities. 2) similarly, 101 has probability $x/2$ of transitioning into either 011 or 110, and probability $1-x$ of staying 101, and reverse transitions have the same probability. 3) The configurations 000 and 111 remain unchanged. Writing the Markov matrix for a trimer in the basis of configurations 
\begin{align*}
\{c_j\}=\{ (000), (001),(010),&(011),(100)\ldots (111) \}\\
&\downarrow\\
\{m\}=\{0,1,2&,3,4,...7\}
\end{align*}
the above transition rules read

\begin{equation}
\label{eq:model}
M=
\begin{bmatrix}
1   &   0&   0&   0&   0&   0&   0&   0\\
0   &   1-y&   y/2&   0&   y/2&   0&   0&   0\\
0   &   y/2&   1-y&   0&   y/2&   0&   0&   0\\
0   &   0&   0&   1-x&   0&   x/2&   x/2&   0\\
0   &   y/2&   y/2&   0&   1-y&   0&   0&   0\\
0   &   0&   0&   x/2&   0&   1-x&   x/2&   0\\
0   &   0&   0&   x/2&   0&   x/2&   1-x&   0\\
0   &   0&   0&   0&   0&   0&   0&   1\\
\end{bmatrix}
\end{equation}

\begin{figure}[h]
    \centering    \includegraphics[width=1\linewidth]{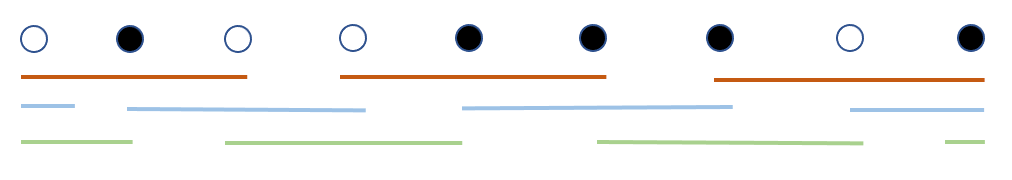}
    \caption{The three triplets represented in red, green and blue colors chosen randomly for updates (random) and sequentially (Floquet) under periodic boundary condition. The filled circles represents a particle(1) and empty circles represents hole (0).}
    \label{fig:update_visualization}
\end{figure}

We maximize the mobility of 1 particle in a trimer (it jumps anywhere in the trimer with probability 1) by setting $y=1$ while controlling the mobility of one hole (2 particle sector) with the parameter $x$. We will consider four values of $x=0, 0.1, 0.5, 1$ and $y=1$ in this section but will only work with $x=0$ and $y = 1$ in the following sections. 

\subsection{Diffusion constant}

In a diffusive system with a single conserved density, the hydrodynamics consists of a continuity equation 

\begin{align}
\label{eq:continuity}
    &\dot{\rho}+\nabla\cdot j = 0,
\end{align}

which implicitly defines the current $j$, and a constitutive relation for the current (i.e., Fick's law), 
\begin{equation}
    j(r,t)=-D \nabla \rho(r,t) +\eta(r,t).\label{eq:fick}
\end{equation}
We take $\eta$ to be Gaussian white noise with a variance related to $D$ by the fluctuation-dissipation theorem. In principle Eq.~\eqref{eq:fick} can contain terms with more derivatives, as well as nonlinear terms that come from the density-dependence of $D$ and $\eta$. The physical diffusion constant (possibly renormalized away from $D$) is defined in the standard way as the growth of the variance of an initially localized density perturbation, i.e.,
\begin{equation}
    \langle r^2\rangle=2 D(t) t,\quad D(t) \equiv \lim_{Q\to 0} \partial^2\langle \rho_Q(t)\rho_Q(0)\rangle/\partial Q^2.
\end{equation}
Note that the correct order of limits is to take $Q \to 0$ first, leaving a time dependent (in principle) diffusion constant, whose late-time limit is related to DC conductivity by the Einstein relation.

\begin{figure}[h]
    \centering
    \includegraphics[width=\columnwidth]{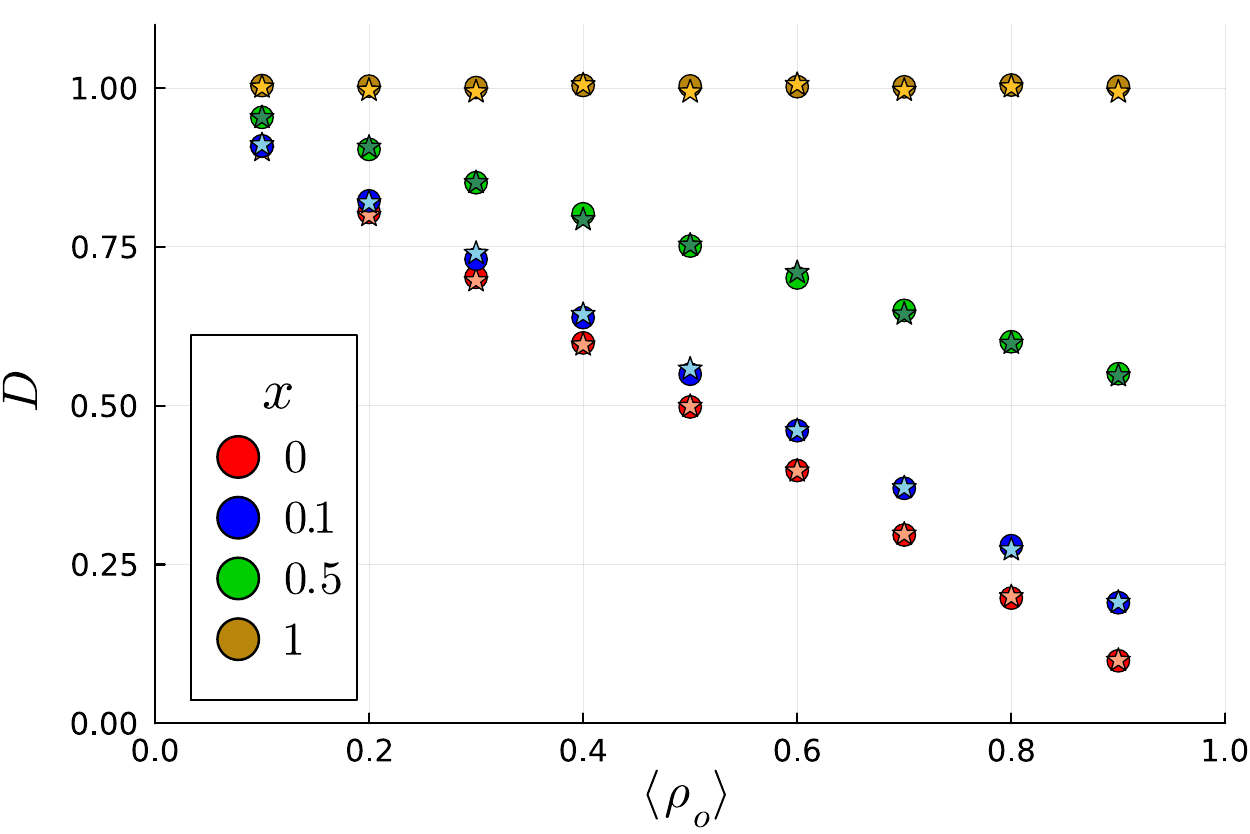}
    \caption{
    Diffusion constant for symmetric(gold) and asymmetric(red, blue and green) update rules for $y=1$, calculated in two different ways as Star(from microscopic current) and Circle(from log($\rho$)) described in the text. We used $Q = \pi/50$ and averaged over $10^5$ samples.
    }
    \label{fig:Dvsn}
\end{figure}

Numerically one can extract the diffusion constant $D$ in Fourier space in two equivalent but operationally different ways:
\begin{enumerate}
    \item by observing exponential decay of density autocorrelation at momentum $Q$ with rate $\Gamma_Q=D Q^2 t$ (as implied by combining  Eqs.\ref{eq:continuity}, \ref{eq:fick} to eliminate current);
    \item by observing both density and current at momentum $Q$ and computing the appropriate ratio as implied by Eq. \ref{eq:fick} (this assumes the noise vanishes on average).
\end{enumerate}
Note, these implicitly require taking thermodynamic $L\to \infty$ limit first. As we explore in Sec. \ref{sec:meso} and a companion paper\cite{Raj_2025} the diffusion constant may be extracted from asymptotic \emph{late time} decay albeit strictly of the lowest allowed Fourier mode $Q=2\pi/L$.

\subsection{Long-time tails}
In general, $D$ depends on the local density. Expanding to leading nontrivial order, Fick's law then becomes $j(r,t) = - D_0 (1 + \alpha \rho(r,t)) \nabla \rho(r,t)$. This nonlinearity is irrelevant at the diffusive fixed point, but (as we noted above) can be dangerously irrelevant, altering the late-time asymptotics of correlation functions. In ``conventional'' long-time tails, physical correlation functions decay as power laws in time. These power laws arise (for example, in current-current correlators~\cite{PhysRevB.73.035113}) at low orders in perturbation theory in the nonlinearity. For the finite-$Q$ response, the perturbative  long-time tails are not kinematically allowed, so the leading allowed process is the nonperturbative diffusion cascade~\footnote{However, if the diffusion constant is random in space, even the finite-$Q$ response is dominated by a power-law long-time tail~\cite{ernst1984long, PhysRevB.29.1755}.}. 

The coefficient $\alpha$ can be determined by numerically extracting the density-dependent diffusion constant (Fig. \ref{fig:Dvsn}). 

\subsection{Validation of nonlinear hydrodynamics using second harmonic generation}
\label{sec:2ndharm}

We perform a perturbative analysis using the equation of motion to get the short time dynamics of different Fourier modes. Here we ignore the noise and its effects on the other modes for short time evolution of small momentum modes.

\begin{figure}[h!]
    \centering
        \includegraphics[width=0.95\columnwidth]{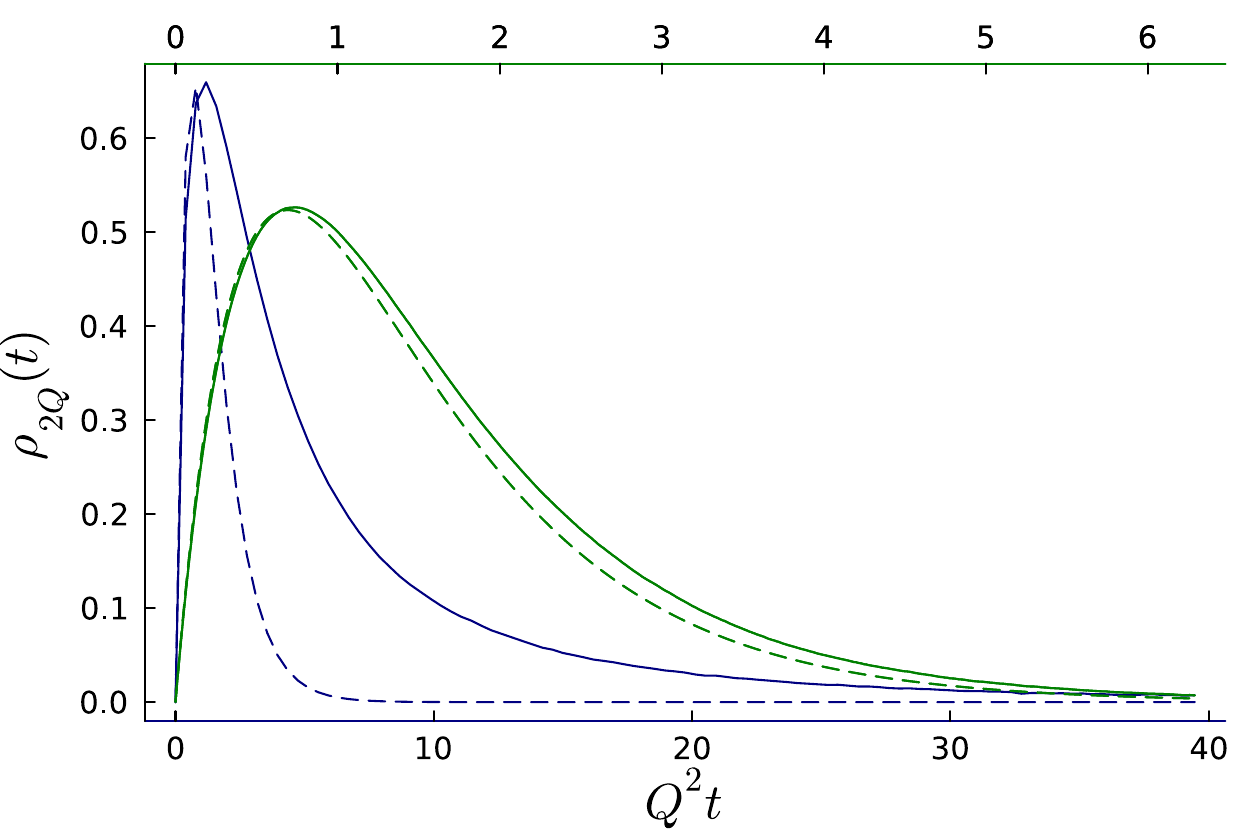}
    \caption{$2^{nd}$ Harmonic Generation for two different Q's(blue:$\pi$/5, green: $\pi$/50) and qualitative comparison with analytical result. $L=99999$ averaged over 1 million samples.}
    \label{fig:2ndHarm}
\end{figure}

\begin{equation}
    \left( \partial_t + D_0 q^2 \right) \rho_{q}(t) + \frac{D_1}{2} q^2 \int_{q'} \,\frac{dq}{2\pi}' \rho_{q'}(t) \rho_{q-q'}(t) = 0
    \label{eq:inte_nld}
\end{equation}

where $D_1= \alpha D_0$. We solve Eq.~\eqref{eq:inte_nld} for the $Q$ and $2Q$ modes only keeping contributions from the $Q$ mode and ignoring noise to get Eqs.\ref{eq:Q mode}, \ref{eq:2nd_harm}.

\begin{gather}
    \rho_Q(t) = \rho_Q(0)e^{-D_0Q^2t} \label{eq:Q mode}\\ 
    \rho_{2Q} = -\frac{D_1 \rho^2_Q(0)}{2\pi D_o} e^{-2D_oQ^2t}\left(1-e^{-2D_oQ^2t}\right)
    \label{eq:2nd_harm}
\end{gather}

 This analytical result for the $2Q$ mode is plotted and compared with the numerical result \ref{fig:2ndHarm}. We can see the agreement between the two at short times. But at late times the signal decays slower than predicted the this short time analysis. The agreement gets better with decreasing momentum of the initially excited mode.

\section{Results}
\label{sec:results}

\subsection{Long-time relaxation in  infinite lattices}
\label{sec:longtimes}

\begin{figure} 
\includegraphics[width=\columnwidth]{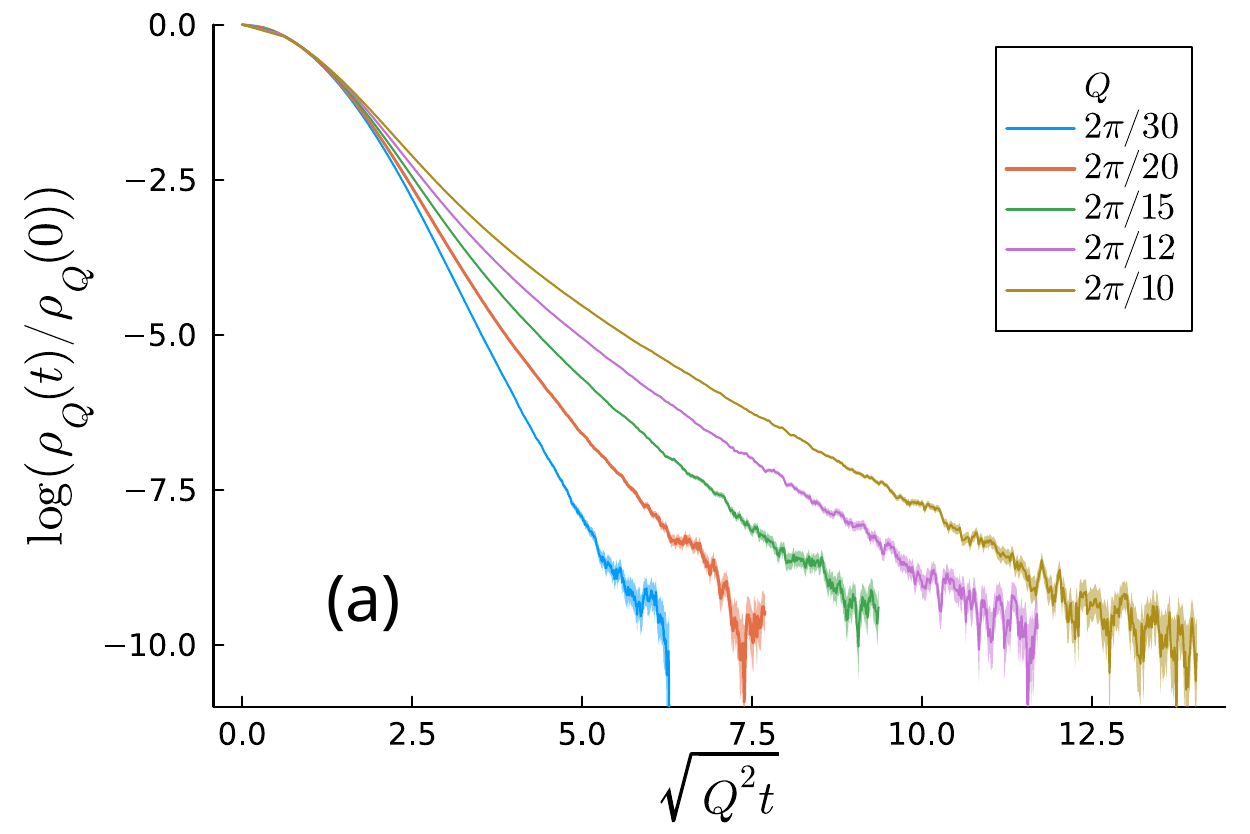} \includegraphics[width=\columnwidth]{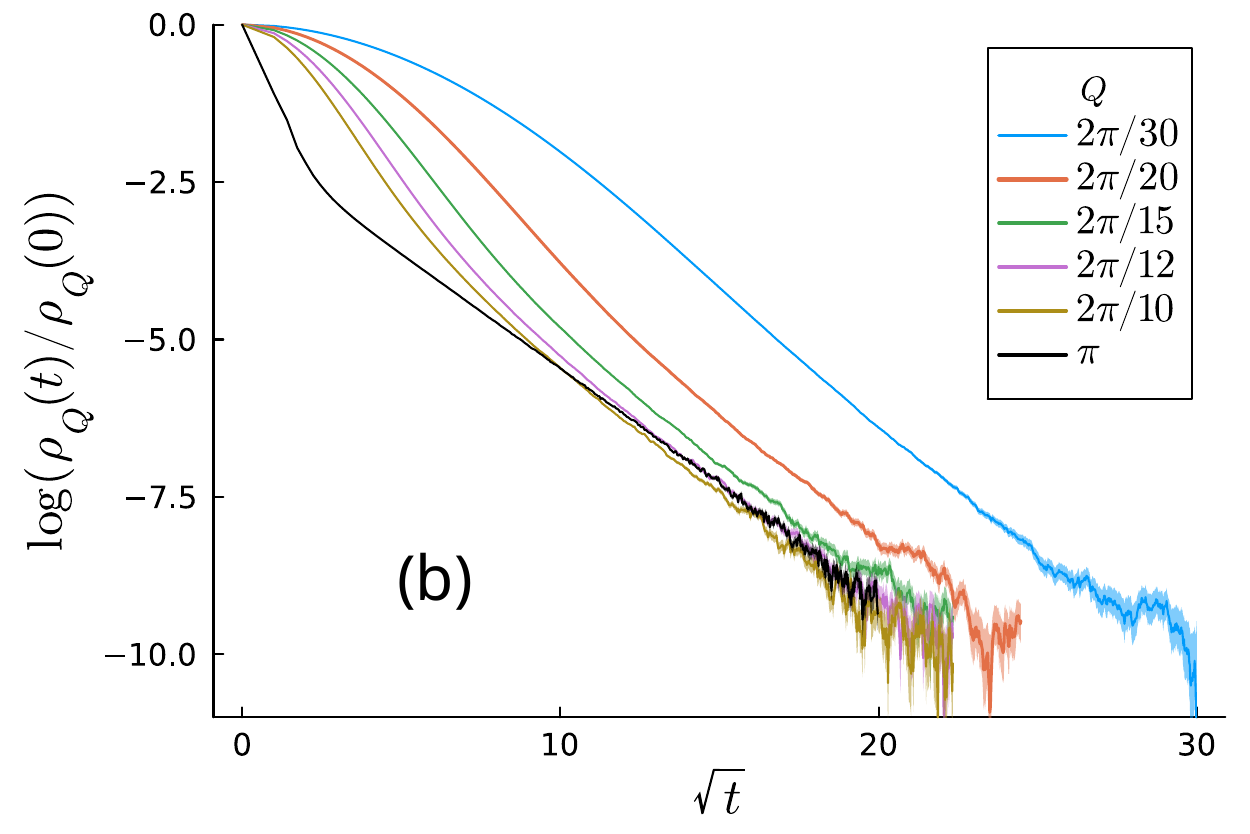}
\includegraphics[width=\columnwidth]{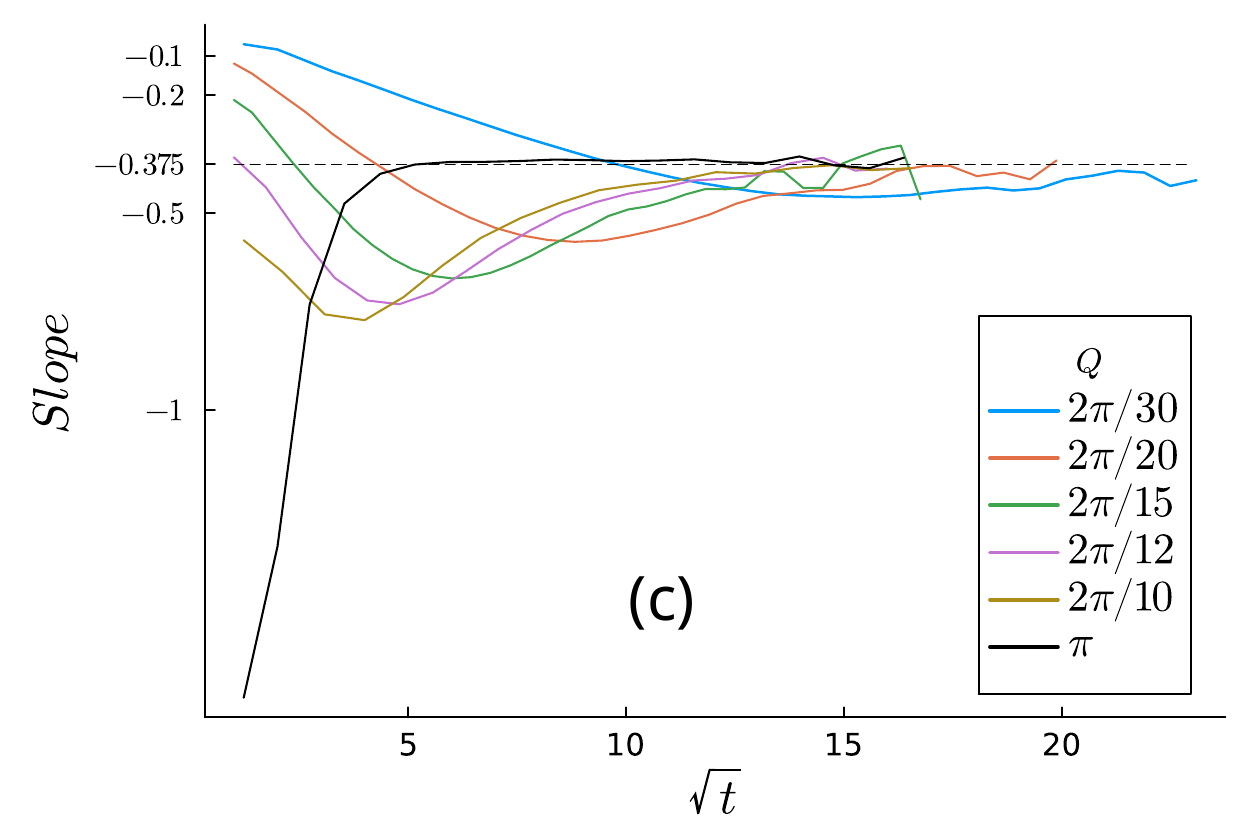}
    \caption{(panel a) Cross-over from diffusive relaxation $\log G(Q,t)\sim - Q^2 t$ at short time to stretched exponential $\log G(Q,t)\sim -\sqrt{t}$ for several Fourier momenta. Note that diffusive $Q^2 t$ rescaling reliably collapses short time data and  is also expected to enter the conjectured stretched-exponential tail at late times\cite{heavyluca}. This is clearly not borne out in the data presented, which is more consistent with a simple $Q$ independent asymptote $\log G(Q,t)\sim -C \sqrt{t}$ (panel b) with $C\approx 0.4$ (panel c).
 This data was obtained with random (non-Floquet) model at $x=0, y=1$ and $L=99999$ averaged over 1 million samples. Note: to obtain good reading on slopes we binned the data on $\sqrt{t}$ axis, ensuring progressively larger bins as you move further to late times i.e. with bin boundaries given by $t_{k}= \left( \frac{k}{m} \cdot \sqrt{t_{max}} \right)^2$ where $t_{max}$ is the total time steps and $m$ is the total number of bins.}
    \label{fig:latetimesRandMain}
\end{figure}

Recall, the central surprising conjecture of LD\cite{heavyluca} is 
\begin{equation}
    \log G(Q,t)\to -\sqrt{Q^2 t} 
    \label{eq:stretchedLD}
\end{equation}
at late times. We now present numerical results that \emph{partially} support this conjecture. 

Fig. \ref{fig:latetimesRandMain} shows plots of $\log G(t)$ vs $\sqrt{Q^2 t}$ for multiple values of $Q$. At short times, the data collapses onto a parabola, corresponding to the prediction from the linear diffusion equation. At late times, each curve crosses over to a linear behavior, consistent with $\log G \sim -\sqrt{t}$. However, the slope of this linear dependence is inconsistent with $|Q|$: instead, empirically, the slope is approximately momentum-independent (Fig. \ref{fig:latetimesRandMain}b) (up to weak transient curvature effects, see Fig. \ref{fig:latetimesRandMain}c). 
However, the crossover from the linear diffusive regime to the nonlinearity-dominated regime seems to be set by $Q^2 t$: it happens at much later time at smaller $Q$. 
These statements can be reconciled if we take $G(Q,t)\to f(Q) e^{-C \sqrt{t}}$, where $f(Q)$ is a rapidly varying prefactor $f(Q\to 0)\to 0 $ but $C\approx 0.4$ has weak momentum dependence.

As expected [Figs. \ref{fig:latetimexpoint2F}], the crossover timescale increases for small nonlinearity.
The results for other choices of the model parameters are similar:  see Figs. \ref{fig:latetimexpoint2F}.

A surprising implication of our empirical functional form $\exp(- C \sqrt{t})$ is that (by dimensional analysis) it requires the existence of a finite length scale that takes the place of $Q^2$ in the scaling relationship with time.

Such a length scale could be in principle induced by the dimensionful nonlinearities such as $D_1$. 

At present, we lack of a theoretical understanding of the origin of this length-scale, and we leave it as an interesting topic for future work.

\subsection{Long-time relaxation in mesoscopic lattices}
\label{sec:meso}
\begin{figure*}
\includegraphics[width=0.3\linewidth]{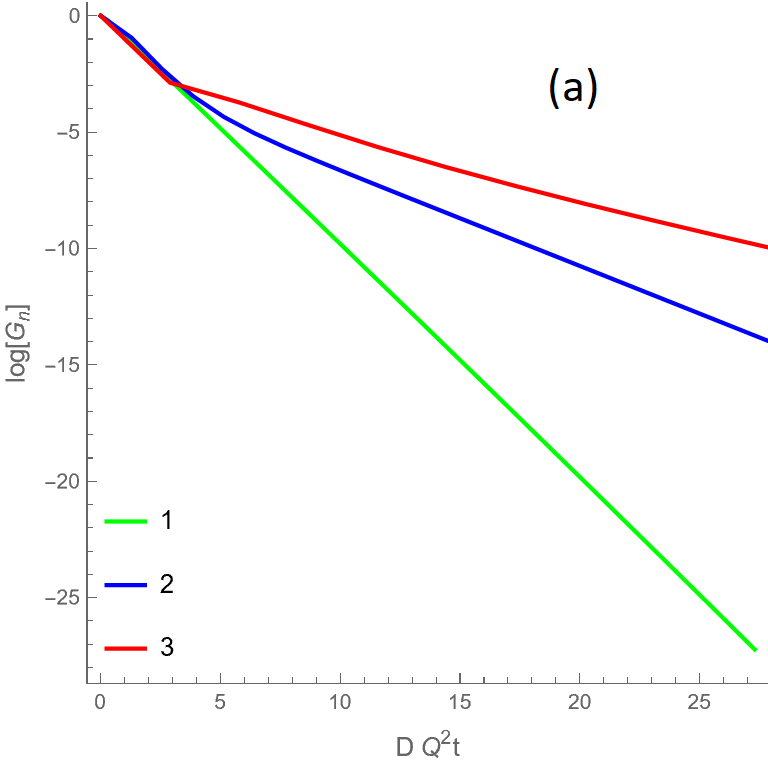}
\hfill
\includegraphics[width=0.3\linewidth]{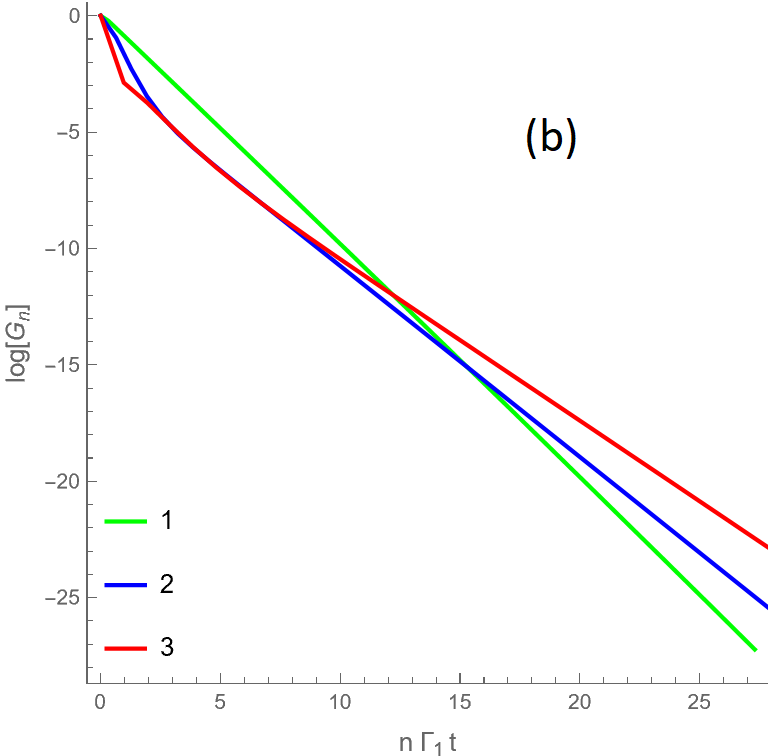}
\hfill
\includegraphics[width=0.3\linewidth]{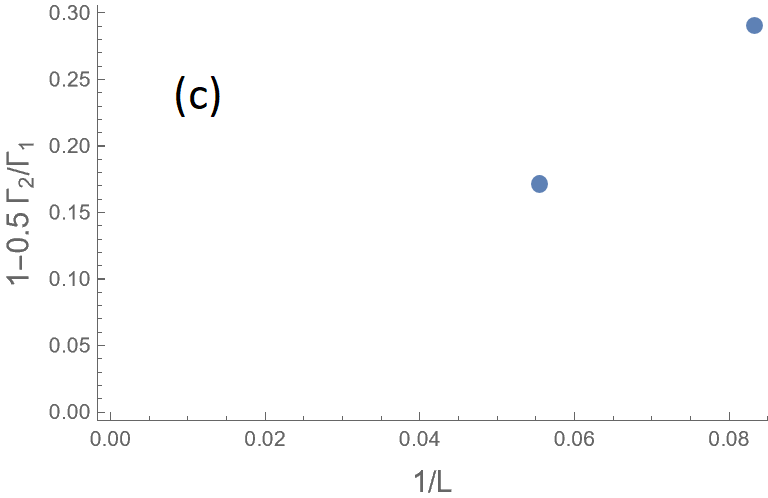}
    \caption{
    Relaxation in small ($L=18$ sites) rings with lowest three modes $Q_n=(2\pi n/L)$ with $n={1,2,3}$. Standard time rescaling $t\to Q_n^2 t$ clearly demonstrates the existence of simple diffusion with diffusion constant $D\approx 2.64$ at short times, while relaxation at late times is governed by a new exponent $\Gamma_n\neq D Q_n^2$ for $n>1$ (panel a). 
    The simplest cascade prediction $\Gamma_n=n \Gamma_1$ does not account for residual interactions among waves which is apparent in our results (panel b): $n^{-1}\Gamma_n/\Gamma_1\approx 0.83, 0.72$ for $n=2,3$, respectively.
    By comparison, similar analysis 
 for $L=12$ (not shown) $n^{-1}\Gamma_n/\Gamma_1\approx 0.71$ for $n=2$. The data from $n=2$ for these two sizes is consistent with $1-n^{-1}\Gamma_n/\Gamma_1\to 0$ as $1/L\to 0$. This deviation is expected (see main text) to vary as $1/L^2$, which is consistent with data shown in panel (c) .
    Note: these results were obtained from exact simulations of 9 particles, i.e. the half-filled sector, using Floquet model with $x=0.1$ -- Secs. \ref{sec:modelsmethods} and \ref{sec:meso} for details.
    }
    \label{fig:meso}
\end{figure*}
Unlike thermodynamic limit, where careful resummation of high order processes is required to know the precise asymptotics, finite size effects offer a simple way to study the cascade order by order in perturbation theory. Simply put, the many-body Fokker-Planck equation is block diagonal in Fourier space, labelled by quantized momenta $Q_n=n 2\pi/L$ and has a gap associated with the slowest mode in each sector. Thus, for our weak quenches we expect a crossover between two exponentials -- with rate $\Gamma_n=D Q_n^2$ at short times and another $n \Gamma_1$ at late times (as was also appreciated in \cite{heavyluca}). Unfortunately, stochastic simulations of dynamics employed till now turned out ill suited to study small systems as they required excessive sampling to reduce the background noise. Using exact noise averaged time evolution in small systems $L=18$ (Fig. \ref{fig:meso}), $L=12$ (not shown) and $L=24$ (forthcoming) we documented a relatively sharp crossover from diffusive $\Gamma_n$ scaling in relaxation rate to approximate  $n \Gamma_1$ ansatz albeit with large, $10\sim20 \%$, corrections consistent with the residual wave interactions due to finite size. The expected $1/L^2$ variation in these residual interactions is consistent with our preliminary data from $L=12, 18$ see Fig. \ref{fig:meso}c.  

Notes on exact simulations: 

\noindent
(i) these exact simulations were performed on a model with weaker nonlinearity, engineered by changing $x=0\to x=0.1$ -- this gives a smaller bare nonlinearity which reduces residual interactions at late times. In the canonical ensemble $x=0$ model has a small number of exactly stuck configurations\cite{Raj_2025} which are not important in the thermodynamic limit but can obscure precise analysis of late time data. This concern was the initial motivation for moving away from $x=0$, as it turns out, simply weaker nonlinearity is the real benefit here -- see item (iii) below; 

\noindent
(ii)
we also switched from random pattern of tripartitions used for large $L$ long-time simulations (see Fig. \ref{fig:update_visualization}) to a fixed  sequence of Markov updates --
this noticeably increases the average diffusion constant and seemingly sharpens the crossover in finite systems; It does not qualitatively change large $L$ and long-time results (Fig. \ref{fig:latetimexpoint2F});

\noindent
(iii) Lastly, and perhaps most importantly, it is crucial to perform exact simulations in the microcanonical ensemble (at half-filling exactly), as density dependence of the diffusion constant produces a mixture of exponents which are visible and confusing to interpret in small systems.

\subsection{Transient signatures of diffusive cascade in wave-coherent multipoint correlators}
\label{sec:cascade}
\subsubsection{Onset and decay of wave coherences in the diffusion cascade}
\begin{figure}[h!]
    \centering
\includegraphics[width=2.6in]{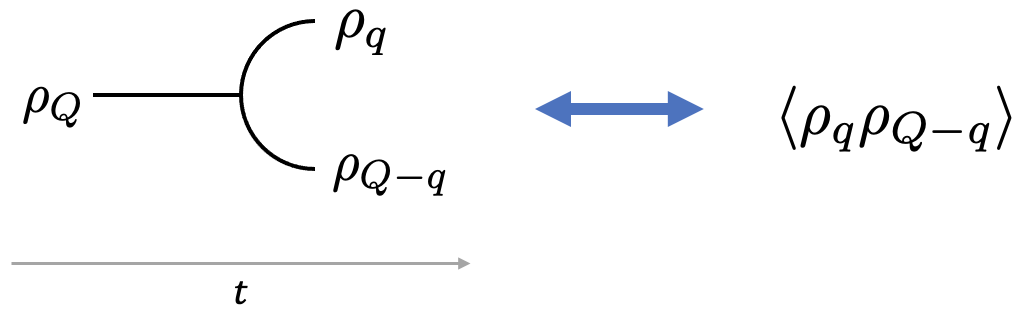}
    \caption{Fourier cascade in real time: initial single wave expectation value at $Q$ seeds a coherent wave pair with the same Fourier momentum.}
    \label{fig:CascadePT}
\end{figure}
\begin{figure}[t]
    \centering
    \includegraphics[width=0.99\linewidth]{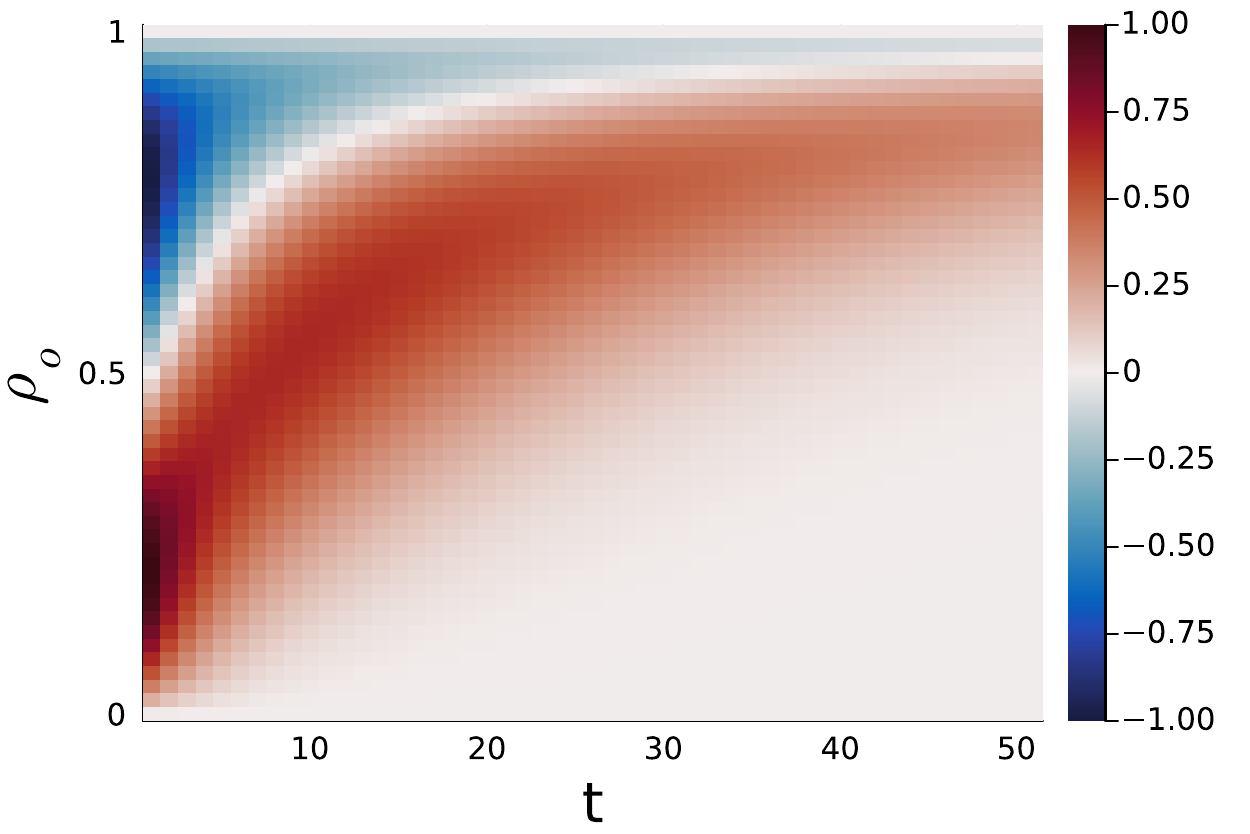}
    \caption{Time- and average denity- dependence of the diffusion cascade eq.\ref{cas-soln} with $Q=\pi/5$ and $q=Q/2$. Normalized with the maximum value.
    }
    \label{fig:heatmap_rho}
\end{figure}

Density dependent diffusion constant is natural\cite{upadhyay2024giant}.
With an eye towards experimental detection of the diffusion cascade in such systems we now explore \emph{coherent} multi-wave observables that are absent in equilibrium but are transiently generated by the cascade. The idea is intuitively clear and we now go through it for the leading perturbative effect(see Fig. \ref{fig:CascadePT}): rather than averaging of intermediate states to calculate perturbative correction to the relaxation of the initially injected wave with Fourier momentum $Q$ we consider cutting the Feynman diagram, i.e. measuring the amplitude of the intermediate two-wave state with Fourier momenta $q, Q-q$. These may be computed analytically (see below, Eqs. \ref{cas-soln} and \ref{cas-soln0}) and also extended to higher orders if necessary. Crucially, these multi-point correlators owe their existence to perfect phase coherence among generated waves, thus any of the individual wave observables with $k\neq Q$ is zero due to random phases, which are nevertheless exactly correlated and lead to finite two-wave amplitudes. This phenomenon is similar to parametric pumping, but differs in that it's entirely off-resonant and transient.  In what follows we explore the details of this coherent wave spectroscopy both as a function of relative momentum $q$ and for the entire density range (away from half-filling), the latter showing additional interesting structure.

\begin{figure*}
    \centering
    \includegraphics[width=0.3\linewidth
    ]{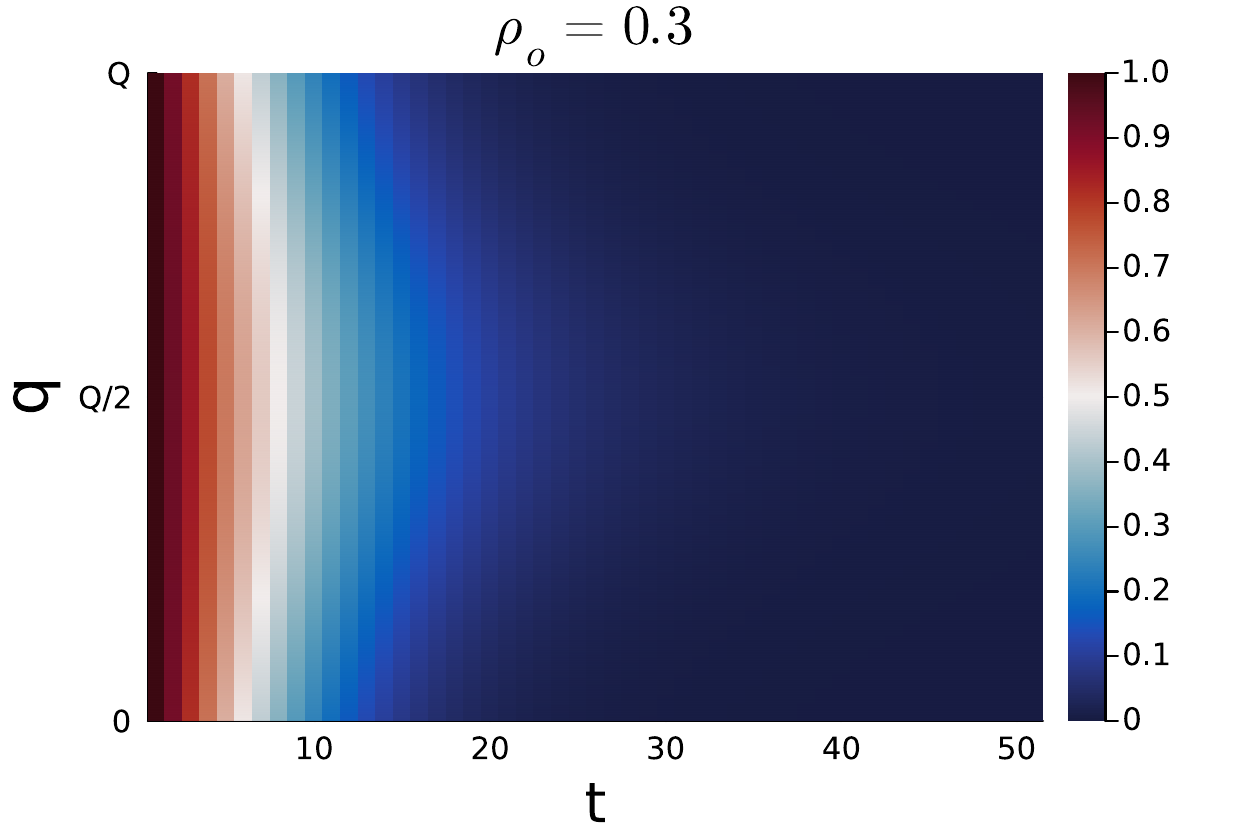}
    \includegraphics[width=0.3\linewidth
    ]{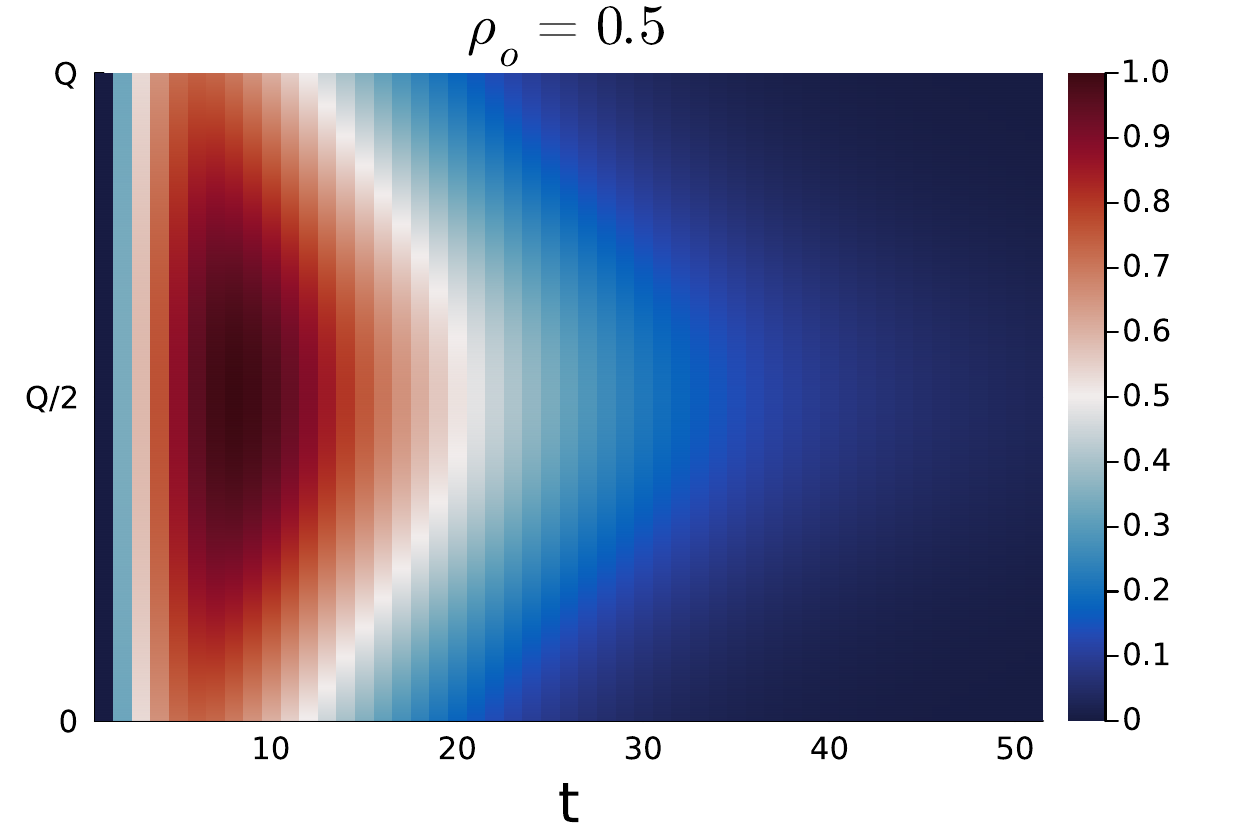}
    \includegraphics[width=0.3\linewidth
    ]{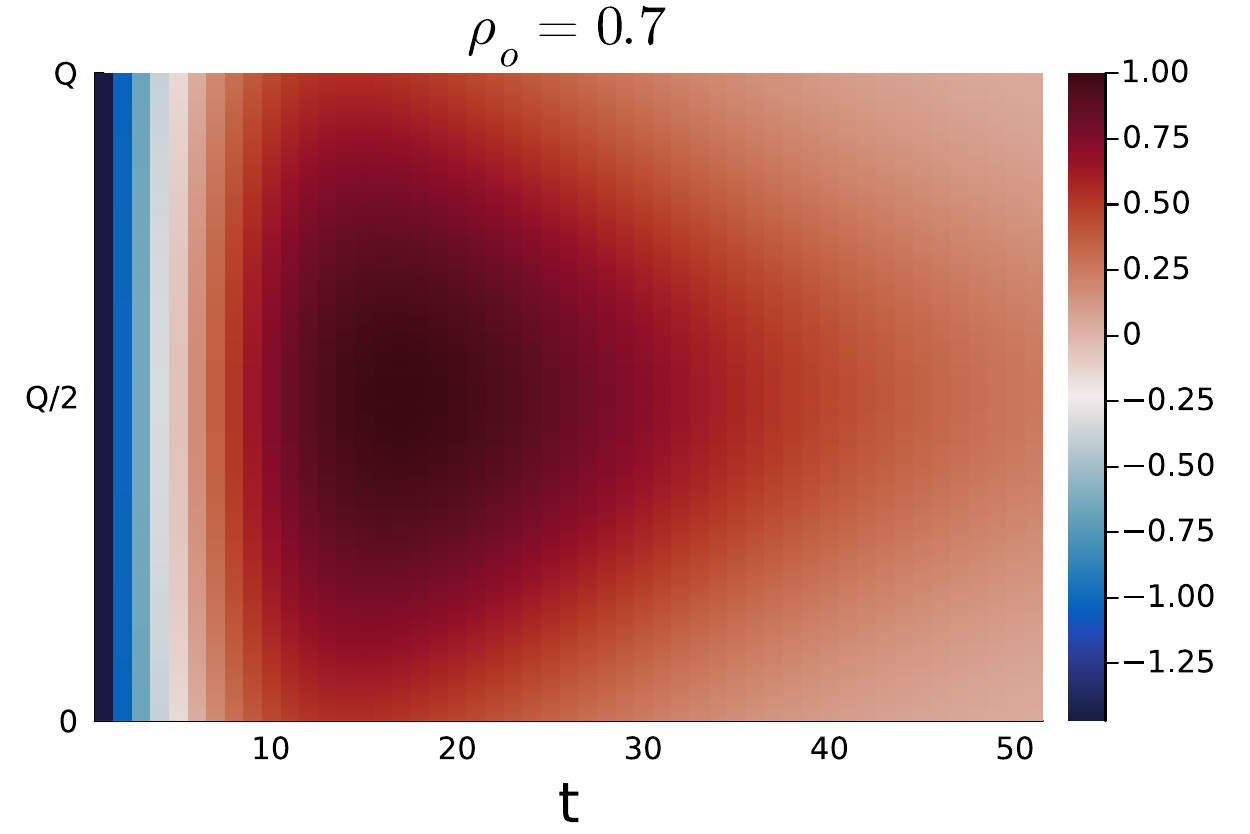}
    \includegraphics[width=0.3\linewidth
    ]{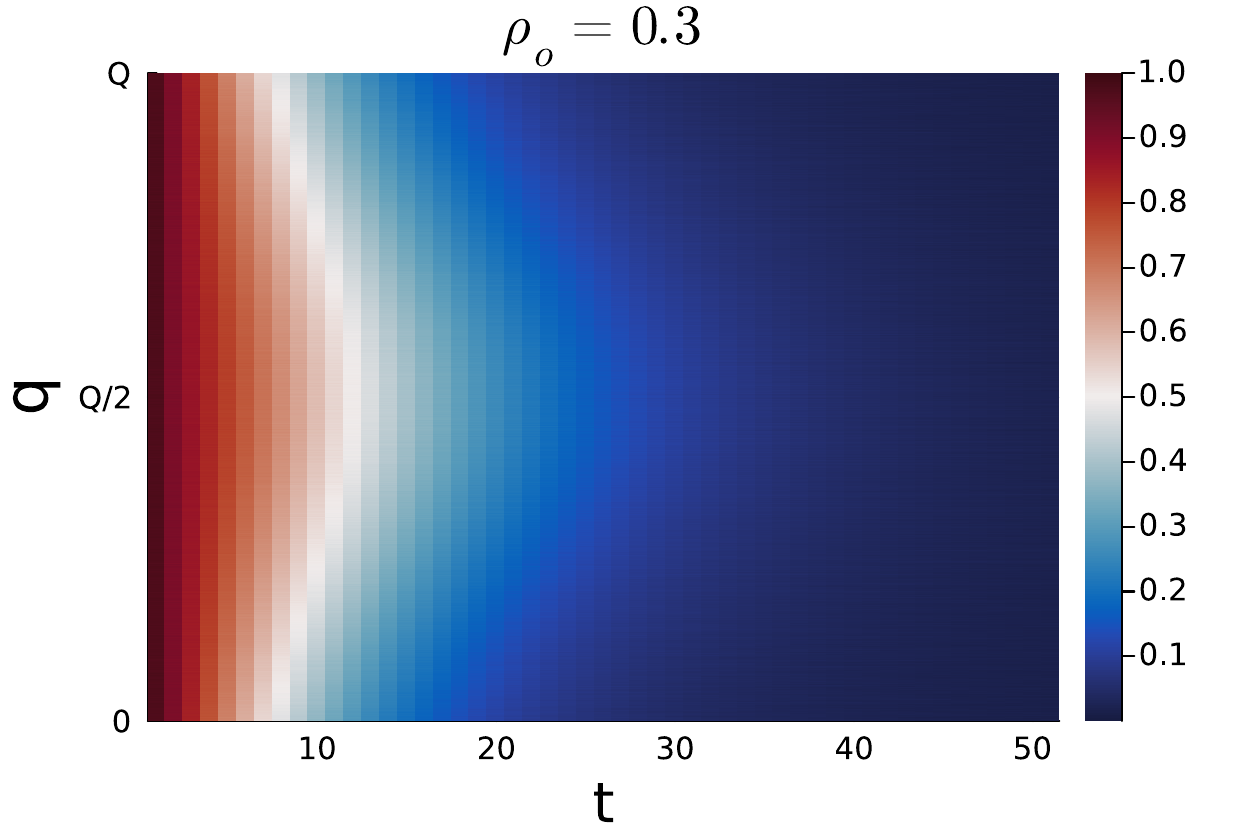}
    \includegraphics[width=0.3\linewidth
    ]{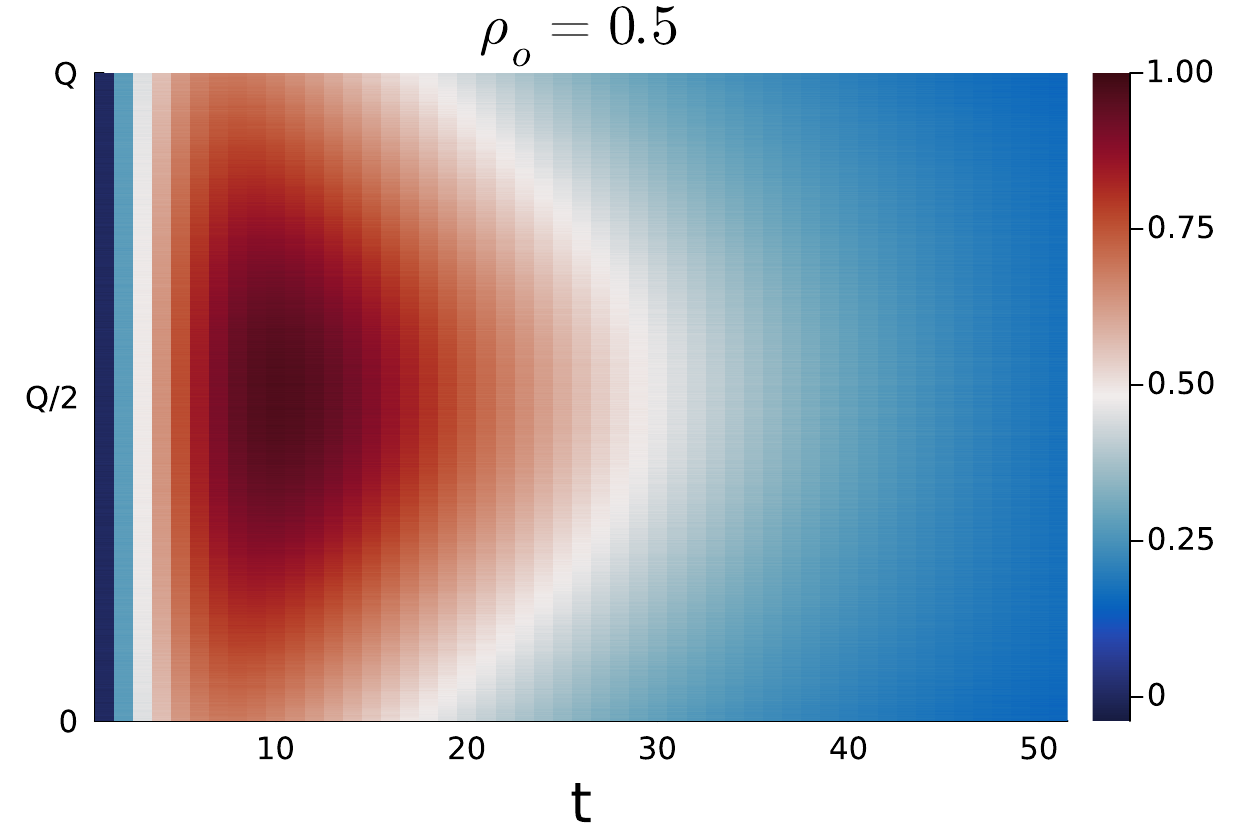}
    \includegraphics[width=0.3\linewidth
    ]{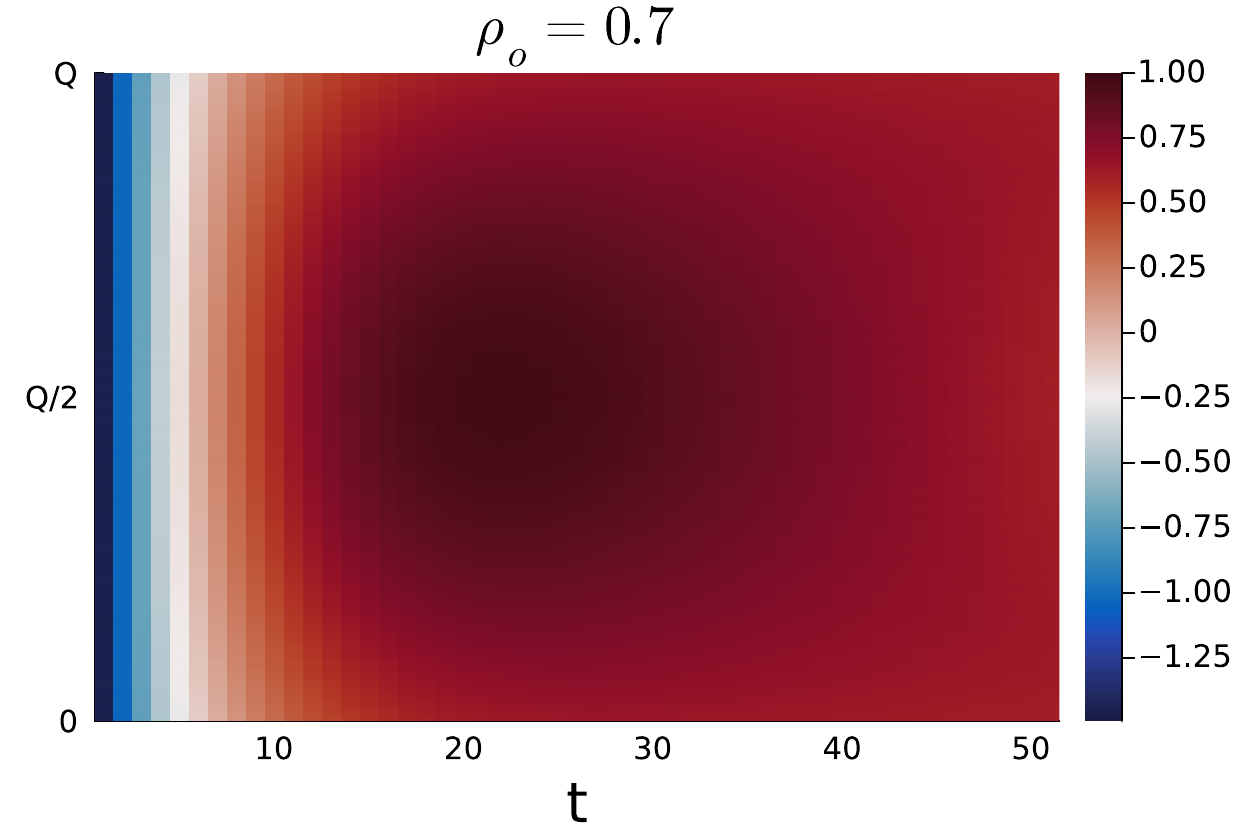}
    \caption{Heatmap of the cascade $\langle \rho_q(t) \rho_{Q-q} (t)\rangle$ vs q and time. Analytical (top) and simulation (bottom) for Q= $\pi$/5 at three different background densities $\rho_o=0.3,0.5$ and $0.7$. To normalize we divide with the maximum value. $L=99999$ averaged over 1 million samples.
}
    \label{fig:cas-theory}
\end{figure*}

To flesh this idea out more we employed the Martin-Siggia-Rose (MSR) formalism to analytically calculate these correlations \ref{fig:CascadePT} and provide insight into the underlying mechanisms driving these long-time behaviors.
The non-linear diffusion equation with the guassian white noise is
\begin{gather}
    \left[\partial_t - D_0 \partial^2_x - D_1 \partial_x (\rho \partial_x ) \right] \rho  + \partial_x \zeta = 0 \\
    \langle \zeta(x,t)\zeta(x',t') \rangle = \sigma \delta(x-x')\delta(t-t') 
\end{gather}

Using MSR formalism these set of equations can be converted to an integral over functionals of noise realizations with action

\begin{multline}
    S[\rho,\varphi_a] = \int_x \int_t \,dx\,dt [i\sigma_0(\partial_x \varphi_a)^2-\varphi_a(\partial_t - D_0 \partial_x^2) \rho \\
    + \varphi_a D_1 \partial_x(\rho\partial_x \rho) + i\sigma_1 \rho(\partial_x \varphi_a)^2]
\end{multline}

where $\sigma = \sigma_o + \rho \sigma_1  $, $ D = \sigma \chi^{-1}$ and $\chi = \rho(1-\rho)$, and $\varphi_a$ is the auxillary variable conjugate to noise.

Given the above action we can calculate the density-density correlations by integrating over the exponential of this action. We can calculate this path integral perturbatively in the interaction strength $D_1$.
We present the result below Eq\ref{cas-soln} and the detailed calculation can be found in Appendix B.


\begin{align}
&\langle \rho_q(t) \rho_{Q-q} (t)\rangle = 
\frac{-\chi}{D_0} e^{-D_0 [q^2+(Q-q)^2]t} \{ [D_1\chi-\sigma_1]+ \nonumber \\
&
\left[ D_1\chi\left( \frac{q^2+(Q-q)^2}{2q(Q-q)}\right) + \sigma_1 \right] (1-e^{-2D_0 q(Q-q)t})\}
\label{cas-soln}
\end{align}

Specializing to half-filling $\chi_1=0$ and $\sigma_1=D_1\chi$
\begin{align}
    & \langle \rho_Q(0) \rho_q(t) \rho_{Q-q} (t)\rangle_{\rho_o=1/2} = \nonumber \\
&    \frac{-2D_1 Q^2 \chi^2}{D_0 q(Q-q)}  e^{-D_0 [q^2+(Q-q)^2]t} \left[1-e^{-2D_0 q(Q-q)t}\right]
\label{cas-soln0}
\end{align}

In Fig.\ref{fig:heatmap_rho}, we present a two-dimensional plot of the solution from Eq.\ref{cas-soln}, illustrating the time evolution of the cascade amplitude as a function of density $\rho$ and time $t$ with $\chi=\rho_o(1-\rho_o)$, $D_0=1-\rho_o$ and $D_1=-1$, where the latter two are extracted from the plot of Fig. \ref{fig:Dvsn} with update rules $(x,y)=(0,1)$. This figure visually captures how the system's dynamics evolve across different equilibrium densities under these update rules. In Fig. \ref{fig:cas-theory}, we compare the theoretical predictions (top panel) with simulation results (bottom panel) for three distinct equilibrium densities $\rho_o$, focusing on the behavior of the system with respect to momentum $q$ and time $t$. These figures demonstrate a qualitative comparison between theory and simulations, highlighting the consistency of the models across varying densities.
\subsubsection{Non-gaussian statistics}
\begin{figure}[h]
    \centering
    \includegraphics[width=\columnwidth]{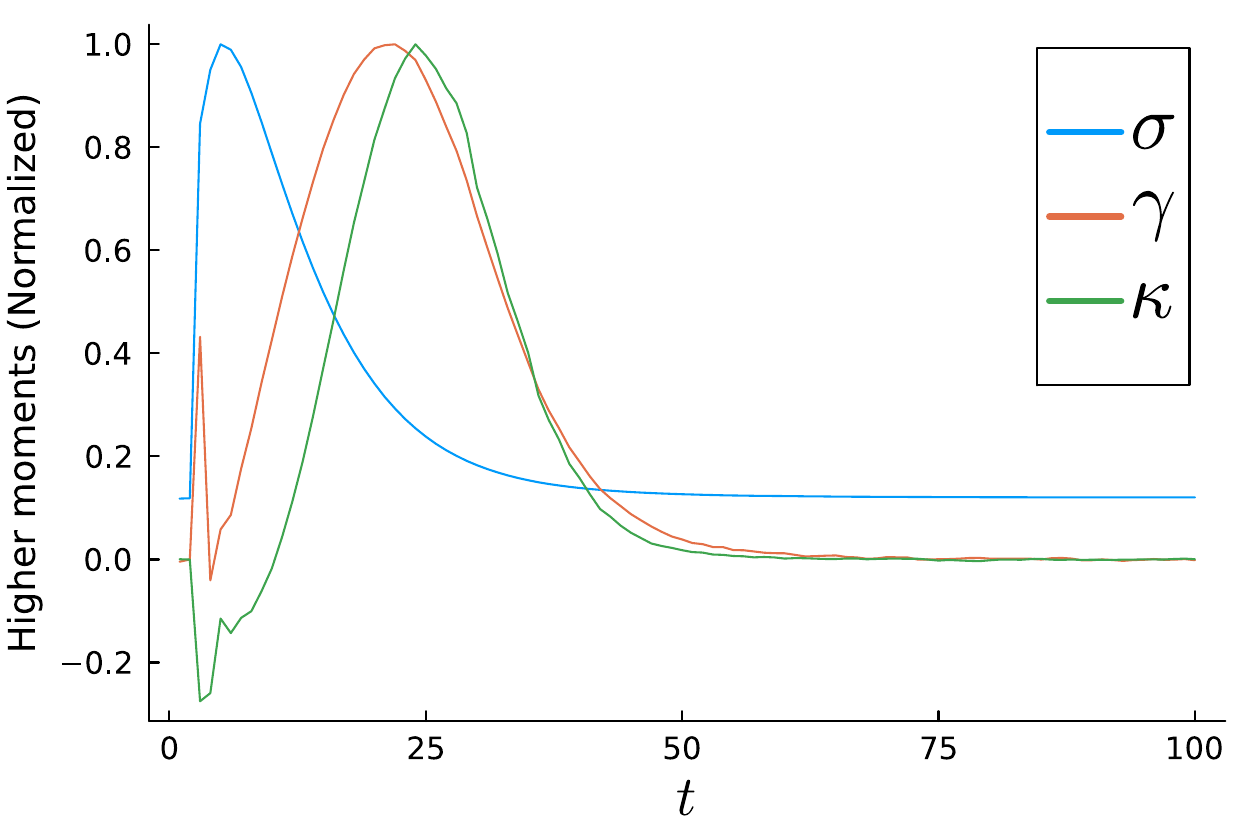}
    \caption{A glimps of nongaussian stats for Q=$\pi$/5, L=99999 and samples = 1000000.}
    \label{fig:beyondmeans}
\end{figure}

Another observable and potentially interesting signature of the diffusive cascade is its (transient) non-gaussianity. Inspired by basic observables used in fluid turbulence we examined variance, skewness and kurtosis of the initially excited wave. Local equlibrium at $t=0$ and $t \to \infty$ is characterized by strictly Gaussian (normal) distribution of $\rho_Q$. Linear diffusion preserves that for all intermediate time, i.e. only the average $A(t)\equiv\langle \rho_Q (t)\rangle$ shows any time dependence as it relaxes exponentially towards equilibrium. The entire distribution remains locally equilibrate (in Fourier space) 
\begin{equation}
    P(\rho_Q)\propto e^{-\chi (\rho_Q-A(t))(\rho_{-Q}-A^*(t))}\Pi_{q\neq Q} e^{-\chi\sum_q \rho_q \rho_{-q}}
\end{equation}By contrast nonlinear diffusion induces non-Gaussian (anomalous) statistics, which we document in Fig. \ref{fig:beyondmeans}. We do not have a theory for this behavior. 

\section{Conclusions}
\label{sec:summary}
In this work we numerically explored the ``diffusion cascade'' proposal~\cite{heavyluca} for the decay of finite-$Q$ perturbations in systems with hydrodynamic nonlinearities. We found that density-wave relaxation is sub-exponential for any fixed $Q$ at sufficiently late times, but the $Q$-dependence seems inconsistent with the predictions of fluctuating hydrodynamics. The density relaxation seems to occur with a $Q$-independent rate, although the crossover to this late-time behavior is set by the diffusive timescale $t \sim 1/Q^2$. On dimensional grounds, this $Q$-independent rate suggests that a length-scale that is invisible at the diffusive fixed point emerges and controls the late time dynamics. 

Motivated by this discrepancy we also checked various other aspects of the diffusion cascade proposal: whether a single wavevector-$Q$ excitation in a finite system decays into a state with multiple excitations at smaller wavevectors, and how the low-order processes that initiate the cascade behave. These results indicate that the basic mechanism by which a finite-$Q$ perturbation decays into a shower of smaller-$Q$ perturbations does in fact take place. In light of these results, the main gap in the argument of Ref.~\cite{heavyluca} seems to be in the treatment of high-order diagrams in perturbation theory. Exploring the diffusion cascade in models that allow for analytical control over the leading diagrams is an interesting direction for future work.

\begin{acknowledgments}
The authors are indebted to Luca Delacr\'etaz, David Huse, Austen Lamacraft, and David Reichman for helpful discussions. We also thank Antonello Scardicchio for collaboration on related work\cite{Raj_2025}.

\end{acknowledgments}

\appendix

\section{Theory of $2^{nd}$ Harmonic Generation}

Here we present the detailed calculation for the $2^{nd}$ harmonic generation. We start with the fourier transform of the noiseless equation.
\begin{equation}
    (\partial_t + D_0 q^2) \rho_{q}(t) + \frac{D_1}{2} q^2 \int_{q'} \,\frac{dq}{2\pi}' \rho_{q'}(t) \rho_{q-q'}(t) = 0
\end{equation}
Then we solve this equation for the momentum mode $Q$ and $2Q$. We get the following equations for both the modes. Where we only keep contributions from the $Q$ mode.
\begin{gather}
    \dot{\rho_Q} + D_0 Q^2 \rho_Q = 0 \\ 
    \dot{\rho}_{2Q} + 4D_0Q^2 \rho_{2Q} + \frac{D_1}{\pi} Q^2 \rho_Q^2(t) = 0
\end{gather}

The first equation for the $Q$ mode simply gives the simple diffusive solution.

\begin{equation}
    \rho_Q(t) = \rho_Q(0)e^{-D_0Q^2t}
\end{equation}

we plug in the solution for this for the $Q$ mode into the equation for the $2Q$ mode to get
\begin{equation}
    \dot{\rho}_{2Q} + 4D_0Q^2 \rho_{2Q} + \frac{D_1}{\pi} Q^2 \rho_Q^2(0) e^{-2D_o Q^2t} = 0
\end{equation}

Now this can be solved by using the solution for the homogeneous equation and multiplying the inverse of it to the non-homogeneous equation to get the exact solution

\begin{gather}
    \dot{\rho}_{2Q} e^{4 D_o Q^2 t} + 4D_0Q^2 \rho_{2Q} e^{4 D_o Q^2 t} + \frac{D_1}{\pi} Q^2 \rho_Q^2(0) e^{2D_o Q^2t} = 0 \\
    \partial_t (\rho_{2Q} e^{4 D_o Q^2 t}) = - \frac{D_1}{\pi} Q^2 \rho_Q^2(0) e^{2D_o Q^2t}
\end{gather}

We integrate both sides to get equation \ref{eq:int}

\begin{gather}
    \int \partial (\rho_{2Q}(t) e^{4 D_o Q^2 t}) = \int_t - \frac{D_1}{\pi} Q^2 \rho_Q^2(0) e^{2D_o Q^2t} \partial t \\
    \rho_{2Q} (t) = -\frac{D_1 \rho^2_Q(0)}{2\pi D_o} e^{-2D_oQ^2t}\left(1-e^{-2D_o Q^2t}\right)
    \label{eq:int}
\end{gather}

\section{MSR Calculation for Diffusion cascade}

We have the non-linear diffusion equation with noise.

\begin{equation}
    [\partial_t - D_0 \partial^2_x] \rho - \frac{D_1}{2} \partial_x^2 \rho^2  + \partial_x \zeta = 0
\end{equation}

with the white noise correlator given by

\begin{equation}
    \langle \zeta(x,t)\zeta(x',t') \rangle = \sigma \delta(x-x')\delta(t-t')
\end{equation}

where $\sigma(\rho) = \chi(\rho) D(\rho)$. And also we can write $\sigma(\rho) = \sigma_o + \rho \sigma_1 $.

The partition function for this equation of motion can be written using the path integral over the noise realization.

\begin{equation}
    \mathcal{Z}[\zeta] = \int \mathcal{D}\rho(x,t) \delta(\dot{\rho} - D_o\partial^2_x \rho - \frac{D_1}{2} \partial_x^2 \rho^2 +  \partial_x \zeta )
\end{equation}

using the integral representation for the delta-function we get another integral over the auxillary field $\varphi_a$ conjugate to the noise field.

\begin{equation}
    \mathcal{Z}[\zeta] = \int \mathcal{D}\rho \int \mathcal{D}\varphi_a e^{-i\int_{xt} dx dt \varphi_a (\dot{\rho} - D_o\partial^2_x \rho - \frac{D_1}{2} \partial_x^2 \rho^2 + \partial_x \zeta)}
\end{equation}

Integrating over the noise with a gaussian distribution

\begin{equation}
    \mathcal{Z}[\rho,\varphi_a] = \int \mathcal{D}\zeta e^{-\int dt dx \zeta^2} Z[\zeta] 
\end{equation}

\begin{equation}
\begin{split}
    \mathcal{Z}[\rho,\varphi_a] = \int \mathcal{D}\rho & \mathcal{D}\varphi_a e^{-i\int dt dx \varphi_a (\dot{\rho} - D_o \partial^2_x \rho - \frac{D_1}{2} \partial_x^2 \rho^2)} \\
    & \int \mathcal{D}[\zeta] e^{-\int dt dx (\zeta^2 - i\varphi_a \partial_x \zeta)}
\end{split}
\end{equation}

The gaussian integral over noise can be performed by summing the squares and we get

\begin{equation}
    \mathcal{Z} = \sqrt{2\pi} \int \mathcal{D}\rho \mathcal{D}\varphi_a e^{\int dt dx [-i\varphi_a(\dot{\rho} - D_o \partial^2_x \rho - \frac{D_1}{2} \partial_x^2 \rho^2) - \sigma(\partial_x \varphi_a)^2]}
\end{equation}

We can read off the lagrangian from the last equation by comparing it to $\mathcal{Z} = \sqrt{2\pi}\int \mathcal{D}\rho \mathcal{D}\varphi_a e^{i\int \,dx \,dt \mathcal{L}[\rho,\varphi_a]}$

\begin{equation}
    \begin{split}
        \mathcal{L}[\rho,\varphi_a] = i\sigma_0(\partial_x \varphi_a)^2-\varphi_a(\partial_t - D_0 \partial_x^2) \rho \\
         + \varphi_a \frac{D_1}{2} (\partial_x^2 \rho^2) + i\sigma_1 \rho(\partial_x \varphi_a)^2
    \end{split}
\end{equation}

We can calculate the expectation value of any observable by doing a path integral over this action

\begin{equation}
    \langle \rho(x,t) \rho(x',t') \rangle = \int \mathcal{D}\rho \mathcal{D}\varphi_a e^{i\int \,dx \,dt \mathcal{L}[\rho, \varphi_a]} \rho(x,t) \rho(x',t')
\end{equation}

To analyse the cascade we want to calculate the contributions of the non linearity to the process depicted in fig. \ref{fig:CascadePT}. There are two distinct types of diagrams contributing to the first order in perturbation.

We perform this this path integral in the fourier space to first order in the interaction.
For the quench setup we have

\begin{gather}
    \langle \rho_q(t) \rho_{Q-q}(t) \rangle_{\mu} = \Tr \left[ \mathcal{P}_o \rho_q(t) \rho_{Q-q}(t)\right] \\
    \mathcal{P}_o = \prod_x \frac{e^{\mu(x)\rho(x,0)}}{1+e^{\mu(x)}} \\
    = \prod_x \frac{1}{2}\left[\left( 1 - \frac{\mu(x)}{2}\right)\delta(\rho(x,0)) + \left( 1 + \frac{\mu(x)}{2}\right)\delta(\rho(x,0) - 1)\right] 
\end{gather}

Expanding to first order in $\mu$

\begin{equation}
\begin{split}
    & \langle \rho_q(t) \rho_{Q-q}(t) \rangle_\mu = \\
    \langle \rho_q(t) \rho_{Q-q}(t) & \rangle_{\mu=0} + \langle \rho_Q(0) \rho_q(t) \rho_{Q-q}(t) \rangle_{\mu=0}
\end{split}
\end{equation}

The first term here is zero for finite Q and q, and calculating the second term gives from last eq to first order in $D_1$ we get.

\begin{equation}
\begin{split}
    & \langle \rho_Q(0) \rho_q(t) \rho_{Q-q}(t) \rangle_{\mu=0} =  \\
    \int_t \mathcal{D}\varphi_a \mathcal{D}\rho \Biggl\langle  \rho_Q(0) \rho_q(t) & \rho_{Q-q}(t)  \left[ \frac{i D_1}{2} \partial_x^2 \varphi_a \rho^2 - \sigma_1  \rho(\partial_x \varphi_a)^2 \right]  \Biggr\rangle
\end{split}
\end{equation}

the two terms here represents the two feynman diagrams drawn below. 

The propagators  for which are defined as

\begin{tikzpicture}
  \begin{feynman}
    \vertex (a1);
    \vertex [right=of a1] (a2);
    \vertex [right=of a2] (a3) ;

    \diagram* {
      (a1) -- [fermion, edge label=\( \rho \)] (a2) -- [fermion, edge label=\( \rho \)] (a3),
    };
  \end{feynman}
\end{tikzpicture}

\begin{equation}
    \mathcal{G}_{\rho \rho}(k,t) = \chi e^{- D_o Q^2 |t|}
\end{equation}

\begin{tikzpicture}
  \begin{feynman}
    \vertex (a1);
    \vertex [right=of a1] (a2);
    \vertex [right=of a2] (a3) ;

    \diagram* {
      (a1) -- [boson, edge label=\( \varphi_a \)] (a2) -- [fermion, edge label=\( \rho \)] (a3),
    };
  \end{feynman}
\end{tikzpicture}

\begin{equation}
    \mathcal{G}_{\varphi_a \rho}(k,t) = - i \theta(t) e^{- D_o Q^2 t} 
\end{equation}

where $\theta(t)$ is Heaviside step function.

For the first diagram 

\begin{tikzpicture}
  \begin{feynman}
    \vertex (a1) {\(\rho_Q\)};
    \vertex [right=of a1] (a2);
    \vertex [right=of a2] (a3) ;
    \vertex [above right=of a3] (b1) ;
    \vertex [below right=of a3] (c1);
    \vertex [right=of b1] (b2) {\(\rho_q\)};
    \vertex [right=of c1] (c2) {\(\rho_{Q-q}\)};

    \diagram* {
      (a1) -- [fermion, edge label=\( Q \)] (a2) -- [boson, edge label'=\(\partial_x^2 \varphi_a\)] (a3),
      (a3) -- [fermion, edge label=\(q\)] (b1) -- [fermion, , edge label=\(q\)] (b2),
      (a3) -- [fermion, edge label=\(Q-q\)] (c1) -- [fermion, , edge label'=\(Q-q\)] (c2),
    };
  \end{feynman}
\end{tikzpicture}

\begin{widetext}
\begin{align}\nonumber
     &=  2 \frac{i D_1}{2} \int dt'  i\chi^2 e^{-D_o [q^2+(Q-q)^2]t}\\
    &\qquad\qquad\qquad\times\left\{[Q^2+q^2+(Q-q)^2]\int_{-\infty}^0  dt' e^{D_o[Q^2+q^2+(Q-q)^2]t'} + [q^2+(Q-q)^2]\int_0^t dt' e^{-D_o[Q^2-q^2-(Q-q)^2]t'}
    \right\} \\
    &= \frac{-D_1\chi^2}{D_o} e^{-D_o [q^2+(Q-q)^2]t}\left[ 1 +  \frac{[q^2+(Q-q)^2]}{2q(Q-q)}\left( 1-e^{-2D_oq(Q-q)t}\right) \right]
\end{align}
\end{widetext}

And for the second diagram

\begin{tikzpicture}
  \begin{feynman}
    \vertex (a1) {\(\rho_Q\)};
    \vertex [right=of a1] (a2);
    \vertex [right=of a2] (a3) ;
    \vertex [above right=of a3] (b1) ;
    \vertex [below right=of a3] (c1);
    \vertex [right=of b1] (b2) {\(\rho_q\)};
    \vertex [right=of c1] (c2) {\(\rho_{Q-q}\)};

    \diagram* {
      (a1) -- [fermion, edge label=\( Q \)] (a2) -- [fermion, edge label'=\(Q\)] (a3),
      (a3) -- [boson, edge label=\(\partial_x^2 \varphi_a\)] (b1) -- [fermion, , edge label=\(q\)] (b2),
      (a3) -- [boson, edge label=\(\partial_x^2 \varphi_a\)] (c1) -- [fermion, , edge label'=\(Q-q\)] (c2),
    };
  \end{feynman}
\end{tikzpicture}

\begin{widetext}
\begin{gather}
    =  - 2 \sigma_1 \int dt'  (-\chi) e^{-D_o [q^2+(Q-q)^2]t}
    \left\{[Q^2+q^2-Qq]\int_{-\infty}^0  dt' e^{D_o[Q^2+q^2+(Q-q)^2]t'} - q(Q-q)\int_0^t dt' e^{-D_o[Q^2-q^2-(Q-q)^2]t'}
    \right\} \\
    = \frac{\sigma_1 \chi}{D_o} e^{-D_o [q^2+(Q-q)^2]t}\left[ 1 -  \left( 1-e^{-2D_oq(Q-q)t}\right) \right] = \frac{\sigma_1 \chi}{D_o} e^{-D_o Q^2 t}
\end{gather}
\end{widetext}

Adding both the terms we get

\begin{widetext}
\begin{equation}
    \langle \rho_q(t) \rho_{Q-q} (t)\rangle_{\mu} = 
    \frac{-\chi}{D_0} e^{-D_0 [q^2+(Q-q)^2]t} \left\{ [D_1\chi-\sigma_1]+
    \left[ D_1\chi\left( \frac{q^2+(Q-q)^2}{2q(Q-q)}\right) + \sigma_1 \right] \left(1-e^{-2D_0 q(Q-q)t}\right)  \right\} 
\end{equation}
\end{widetext}

And at half filling $\chi_1=0$ and $\sigma_1=D_1\chi$, the equation above becomes

\begin{equation}
\begin{split}
    &\langle \rho_q(t) \rho_{Q-q} (t)\rangle_{\mu} = \\
    \frac{-2D_1 Q^2 \chi^2}{D_0 q(Q-q)} & e^{-D_0 [q^2+(Q-q)^2]t} [1-e^{-2D_0 q(Q-q)t}]
\end{split}
\end{equation}
\begin{figure*}
\includegraphics[width=0.6\columnwidth]{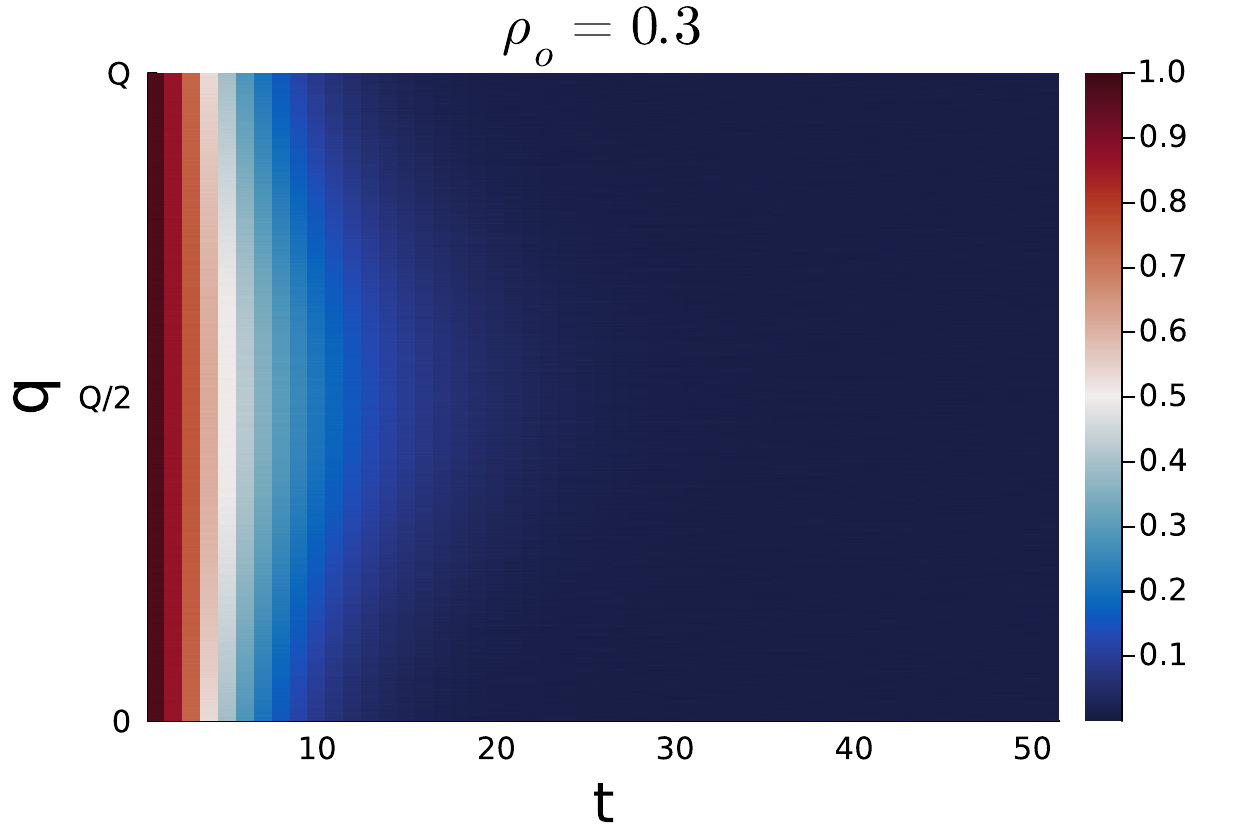}
\includegraphics[width=0.6\columnwidth]{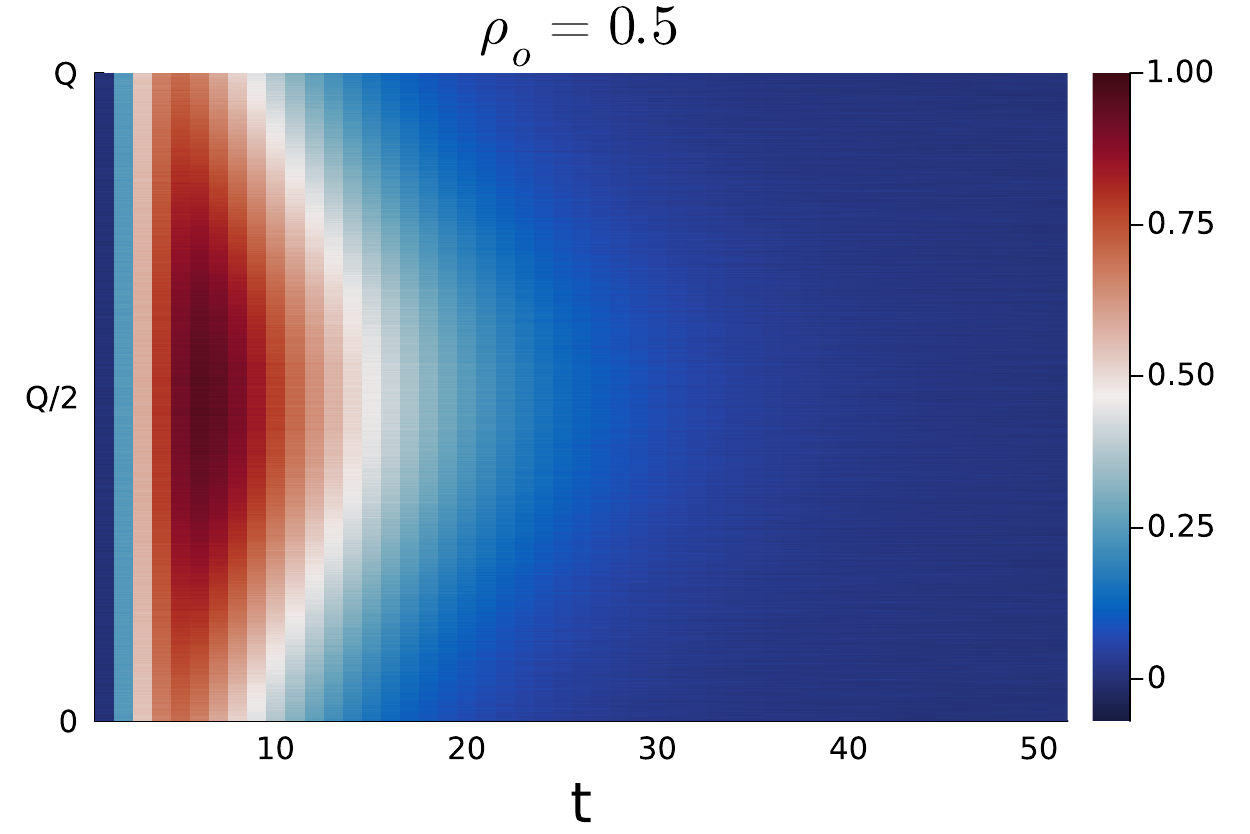}
\includegraphics[width=0.6\columnwidth]{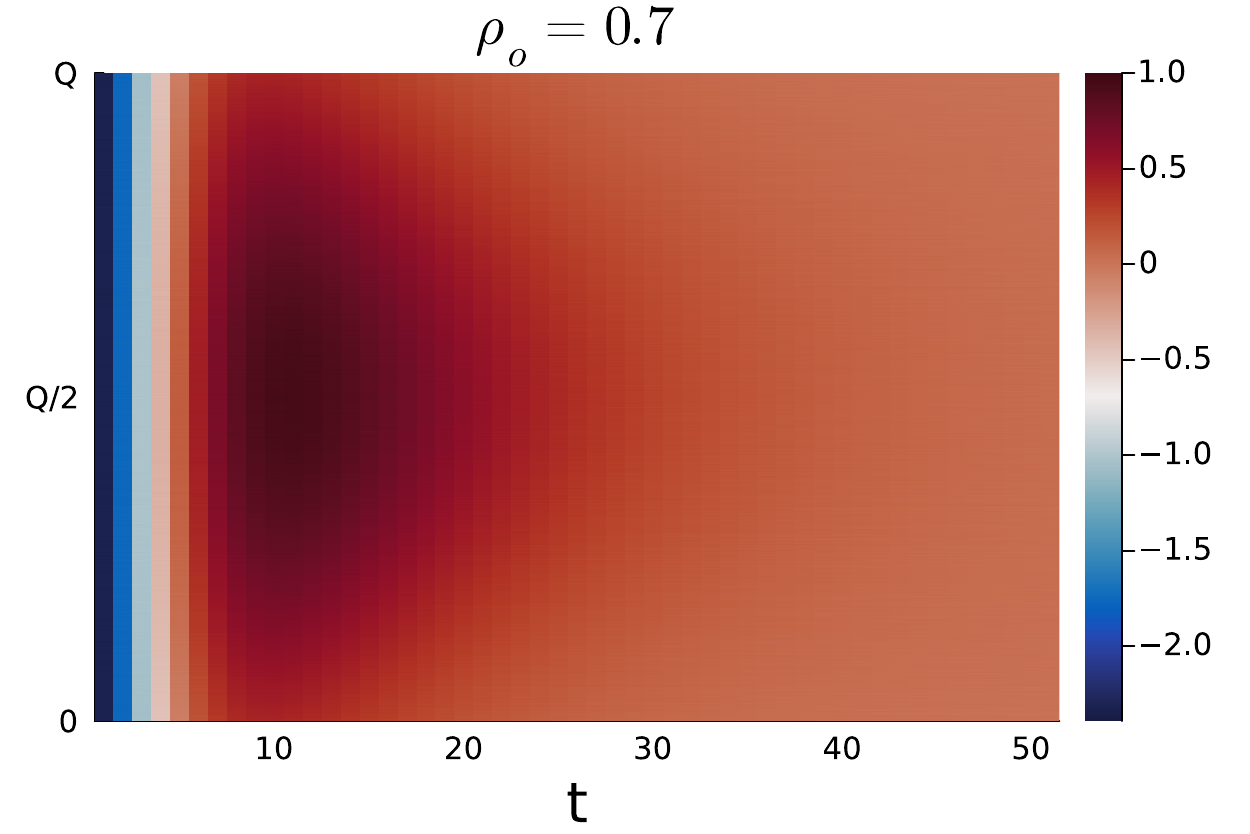}
    \caption{Cascade plots for update rules $(x, y)=(0.2,0.8)$.  Floquet updates}
    \label{fig:casc_new}
\end{figure*}


\section{Additional results and analysis of long-time asymptotics}

Having defined a continuous family of models characterized by two mobilities of single and two particle trimers, $y$ and $x$, respectively, we mostly focused on the extreme limit of $y=1$ and $x=0$, which maximizes the nonlinearity and makes observation of true late time asymptotics easier numerically. As it turns out these extremely strongly interacting models have their own host of additional interesting features\cite{Raj_2025} beyond generic nonlinear hydrodynamics of interest here. To allay possible concerns about interplay of those features and the diffusion cascade we have examined the latter for other, less strongly interacting members of the family, both with random and Floquet patter of updates in this Appendix. The results are consistent with non-dispersive $\sqrt{t}$ relaxation reported in main text.

\begin{figure*}[b!]
\includegraphics[width=0.9\columnwidth]{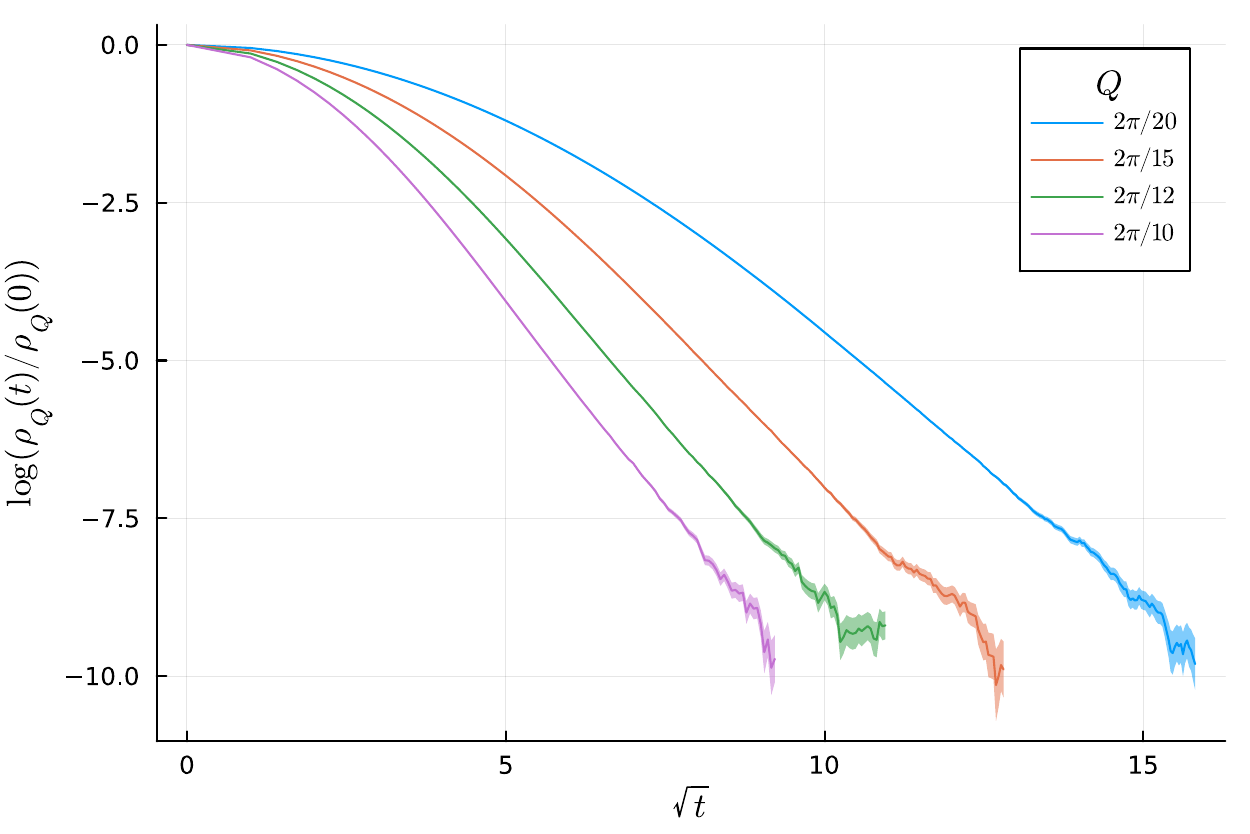}
\includegraphics[width=0.9\columnwidth]{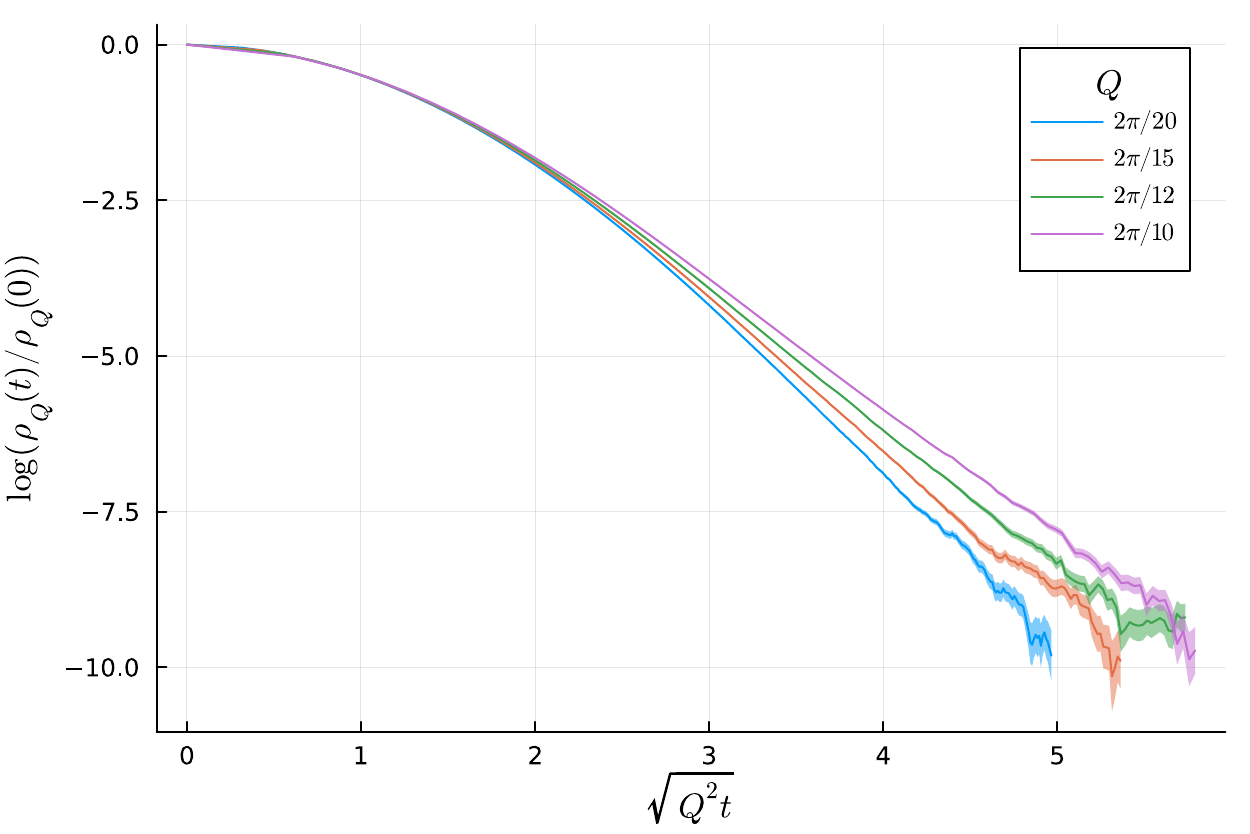}
\includegraphics[width=0.9\columnwidth]{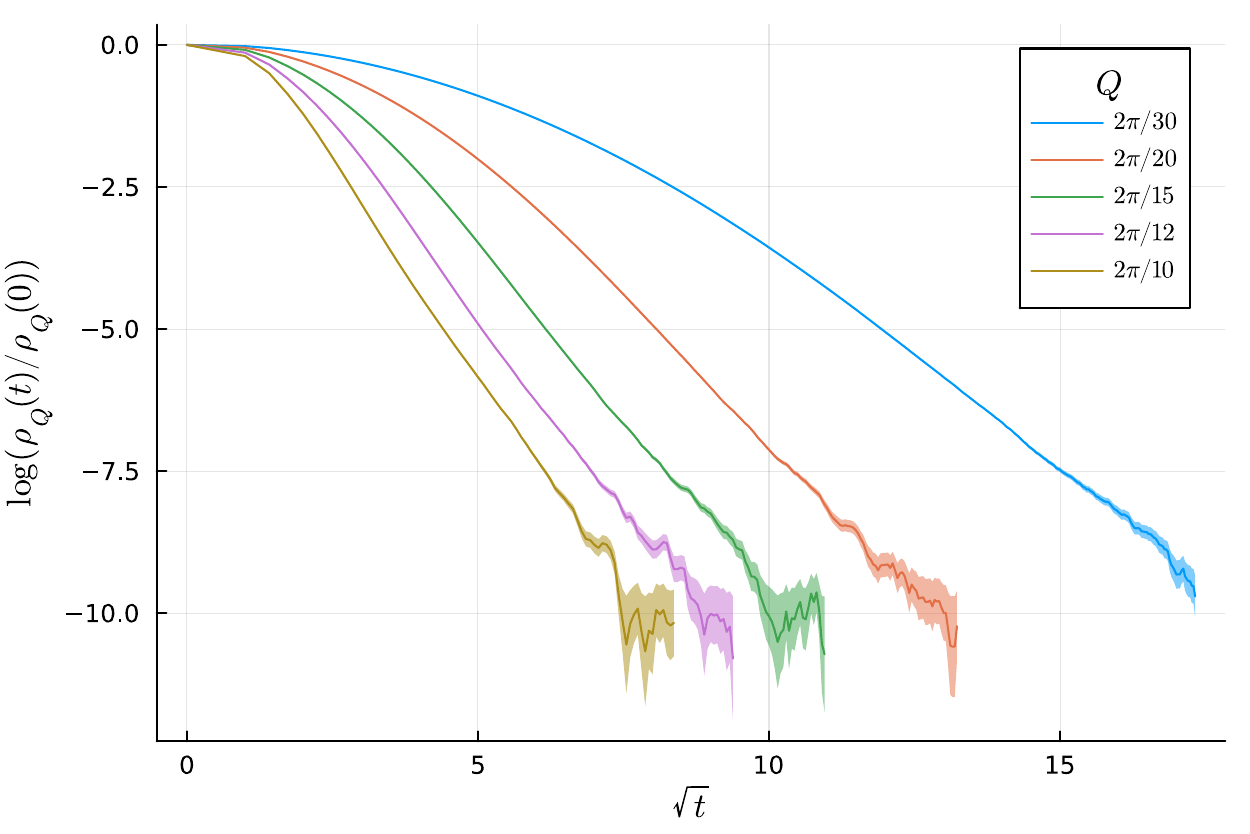}
\includegraphics[width=0.9\columnwidth]{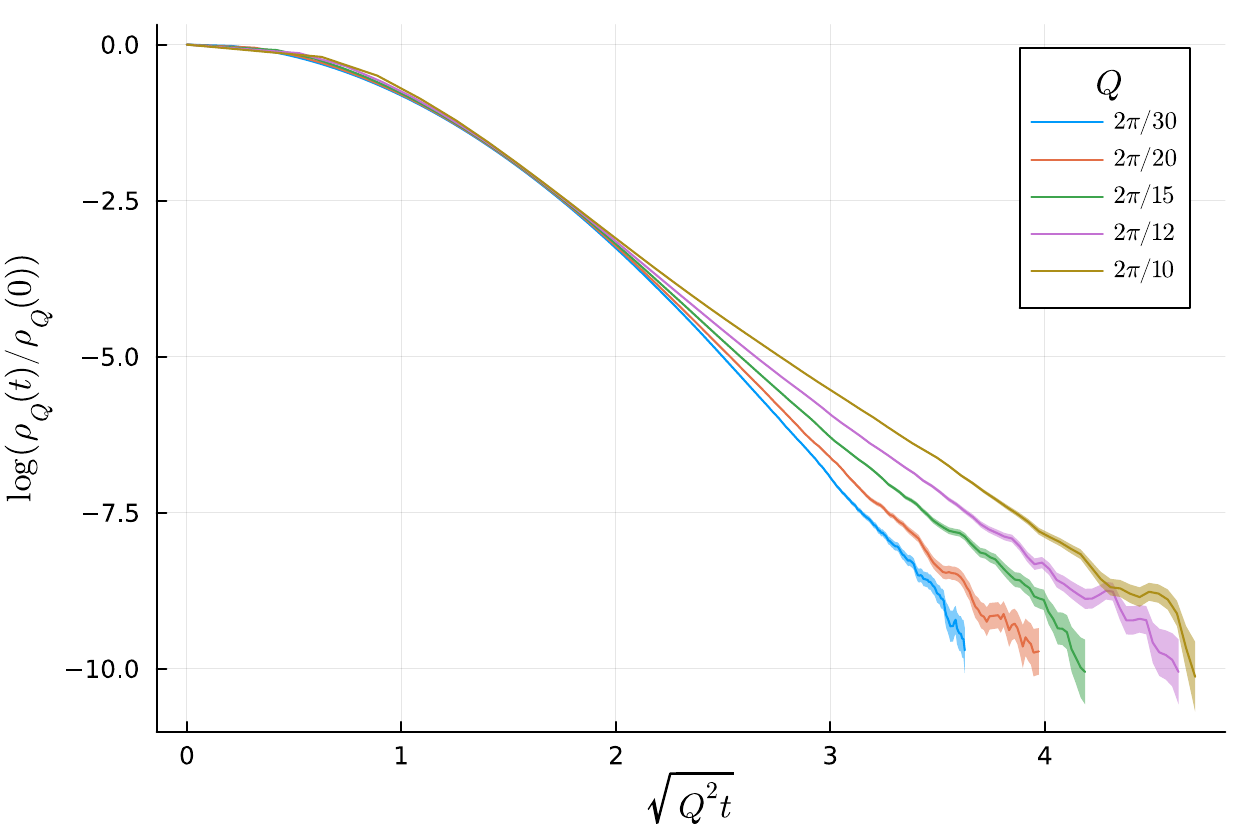}
    \caption{Late-time dynamics in large lattices to compare with Fig. \ref{fig:latetimesRandMain}. Upper panel shows results for x=0.2, y=0.8 with random updates;
Lower panel shows results for x=0.2, y=0.8 with Floquet updates;
}
    \label{fig:latetimexpoint2F}
\end{figure*}

\begin{figure}
\includegraphics[width=0.9\columnwidth]{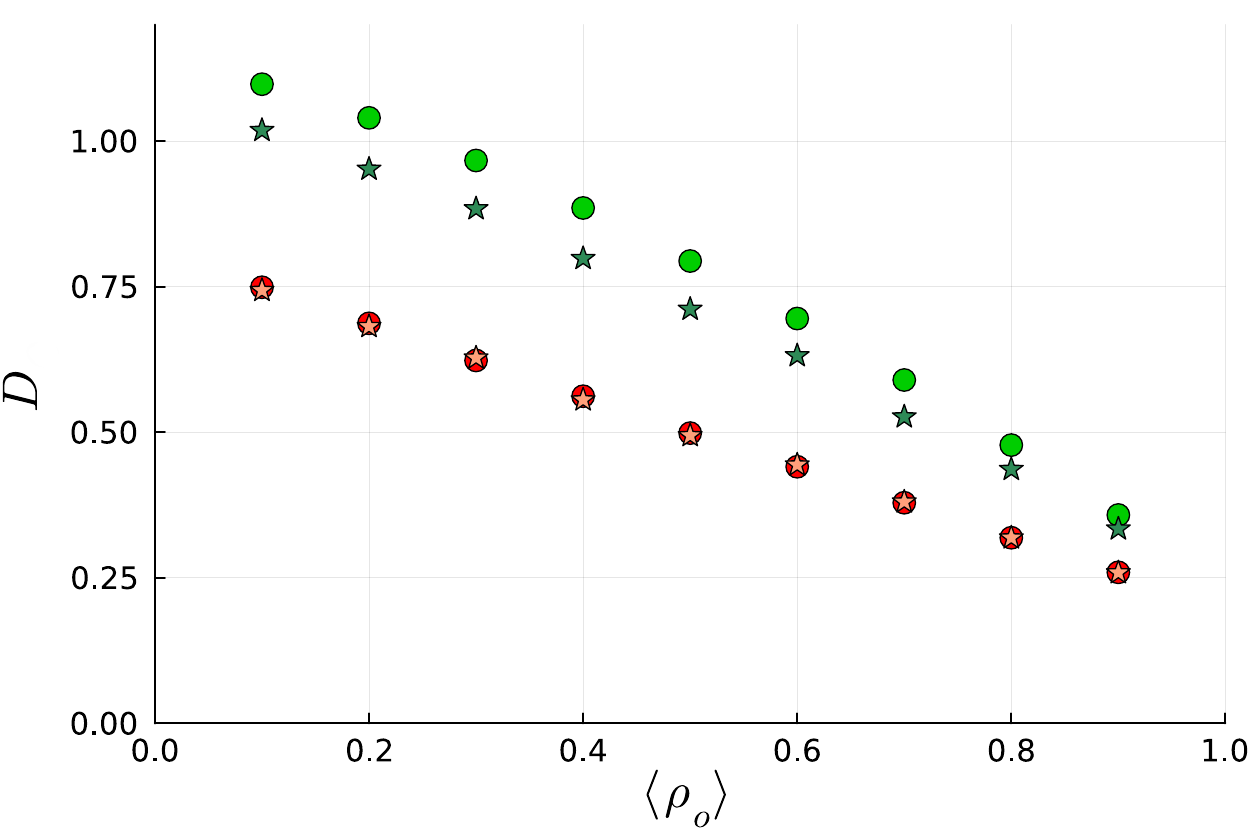}
    \caption{D vs $\rho$ for x=0.2, y=0.8. red $\Rightarrow$ random updates, green $\Rightarrow$ non-random floquet updates. Circles and  stars denote the diffusion coefficient calculated using exponential decay of $Log(\rho)$ and microscopic current respectively.}
    \label{fig:Thouless3}
\end{figure}

\bibliography{Draft}

\end{document}